\documentclass{llncs}
\usepackage[T1]{fontenc}

\usepackage{graphicx}
\usepackage{amsmath}
\usepackage{amsfonts}
\usepackage{mathrsfs}
\usepackage{cleveref}
\usepackage{algorithm,algpseudocode,color}
\usepackage{url}
\usepackage{soul}
\usepackage{subcaption}
\usepackage{wrapfig}
\usepackage{xspace}
\usepackage[dvipsnames]{xcolor}
\usepackage{enumitem}
\usepackage{bbding}

\usepackage{tikz,ifthen,calc}
\usetikzlibrary{positioning, fit}
\usetikzlibrary{arrows.meta}
\usetikzlibrary{positioning}
\usetikzlibrary{shapes}
\usetikzlibrary{shapes.symbols,patterns}
\usetikzlibrary{calc,through,backgrounds}
\usetikzlibrary{decorations.pathreplacing,calligraphy}
\usetikzlibrary{plotmarks}

\newcommand{\set}[1]{\left\{ #1 \right\}}

\newcommand{\calS}{\mathscr{S}}
\newcommand{\calF}{\mathscr{F}}
\newcommand{\IXD}{I_D}
\newcommand{\IYD}{J_D}
\newcommand{\IXS}{I_S}
\newcommand{\IYS}{J_S}
\newcommand{\lowhourglass}{\mathbin{\rotatebox[origin=c]{-90}{$\bowtie$}}}
\newcommand{\hourglass}{\mathrel{\raisebox{1pt}{$\lowhourglass$}}}

\def\FLOW{\mbox{\sf flow}}

\def\tV{{\tilde V}}
\def\tE{{\tilde E}}

\def\inline#1:{\par\vskip 3pt\noindent{\bf #1:}\hskip 10pt}
\def\midinline#1:{\par\noindent{\bf #1:}\hskip 10pt}
\def\dnsinline#1:{\par\vskip -7pt\noindent{\bf #1:}\hskip 10pt}

\newcommand{\DRlong}{\textsc{Degree Realization}\xspace}
\newcommand{\DR}{\textsc{DegR}\xspace}
\newcommand{\BDRlong}{\textsc{Bi-sequence Degree Realization}\xspace}
\newcommand{\BDR}{\textsc{BDR}\xspace}

\newcommand{\domsetlong}{\textsc{Min Dominating Set}\xspace}
\newcommand{\domset}{\textsc{MDS}\xspace}
\newcommand{\match}{\textsc{Max Matching}\xspace}
\newcommand{\vc}{\textsc{Min Vertex Cover}\xspace}
\newcommand{\clique}{\textsc{Max Clique}\xspace}
\newcommand{\biclique}{\textsc{Max Bi-clique}\xspace}
\newcommand{\is}{\textsc{Max Independent Set}\xspace}

\begin{document}
\title{Bi-sequence Degree Realization \\
with a Minimum Dominating Set%
\thanks{Supported by US-Israel BSF grant \#2022205.}
}

\author{%
Amotz Bar-Noy\inst{1}\orcidID{0009-0004-7021-9072}
\and
Igor Kalinichev\inst{2}\orcidID{0009-0009-1609-448X} \Envelope
\and
David Peleg\inst{2} \orcidID{0000-0003-1590-0506}
\and
Dror Rawitz\inst{3}\orcidID{0000-0003-0323-6097}
}

\authorrunning{A. Bar-Noy et al.}

\institute{City University of New York (CUNY), New York, USA.
\email{amotz@sci.brooklyn.cuny.edu}
\and
Weizmann Institute of Science, Rehovot, Israel.
\email{\{igor.kalinichev,david.peleg\}@weizmann.ac.il}
\and
Bar Ilan University, Ramat-Gan, Israel.
\email{dror.rawitz@biu.ac.il}
}

\maketitle              %
\begin{abstract}
The input to the \BDRlong problem is a pair $a=(a_1, \ldots, a_n)$ and $b = (b_1, \ldots, b_m)$ of two integer sequences 
(or a bi-sequence), and the goal is to find a bipartite graph $G =(V,W,E)$, 
such that the degree sequences of $V$ and $W$ are $a$ and $b$, respectively,
or to decide that such a realization does not exist.  
There are well-known polynomial time algorithms for this problem, and there exists a characterization, based on a set of $O(n)$ inequalities that can be used to determine whether a solution exists for a given bi-sequence. 
A bi-sequence $(a,b)$ may have several realizations, and some of them 
may be preferable to others with respect to some measure.
In this paper, focusing on the measure of \emph{dominating set size}, 
we study the problem of \BDRlong with respect to the \domsetlong problem, 
where the goal is to compute a realization of the input bi-sequence $(a,b)$ 
with the smallest possible dominating set among all the realizations of $(a,b)$.
Our first contribution is a polynomial-time realization algorithm for \BDRlong with respect to \domsetlong. 
We also provide a characterization for the existence of a realization 
with a dominating set of size $\gamma$ consisting of $O(n m(n+m)\gamma)$ inequalities.
\end{abstract}

\section{Introduction}

The input in the \DRlong (\DR) problem is a non-increasing sequence 
$d = (d_1,\ldots,d_n)$ of positive integers, 
and the goal is to decide whether there exists a simple graph $G = (V,E)$ such that $d$ is its degree sequence, 
that is, such that $\deg_G(i) = d_i$ for every $i \in V$, and to construct such a graph if it exists. 
A sequence admitting a realization is called \emph{realizable} or \emph{graphic}. 
Erd\H{o}s and Gallai~\cite{EG60} provided a set of $n$ inequalities that determine whether a sequence 
is realizable or not, and this characterization implies an $O(n)$-time decision algorithm.
Havel and Hakimi~\cite{havel55-long,Hakimi62} provided an algorithm that finds a realization for a given 
sequence whose running time is linear in $\sum_i d_i$.

The \BDRlong (\BDR) problem is a variant of \DR in which the input is a pair $(a,b)$, 
where $a = (a_1,\ldots,a_n)$ and $b = (b_1,\ldots,b_m)$,
of two positive integer non-increasing sequences or a \emph{bi-sequence}, 
and the goal is to find a simple bipartite graph $(V,W,E)$, 
such that $|V|=n$, $|W|=m$ and the degree 
sequences of $V$ and $W$ are exactly $a$ and $b$, respectively, 
or to decide that such a realization does not exist.  
The history of \BDR is as long as that of \DR. 
Gale~\cite{Gale57} and Ryser~\cite{Ryser57} gave a set of $n$ (or $m$)
inequalities that determine whether a given bi-sequence is realizable
even before Erd\H{o}s and Gallai's characterization of graphic sequences. 
As for realization algorithms, 
it is well-known that a realization can be found using \textsc{Maximum Flow}.
In addition, there is a folklore realization algorithm for \BDR which is 
similar to the Havel-Hakimi realization algorithm for \DR.

A realizable bi-sequence $(a,b)$ may have several realizations, and in \BDR 
one is asked to obtain any one of them. However, some realizations may be better 
than others. More specifically, given a measure that assigns a score to each 
realization of the bi-sequence, the goal is to obtain a realization with an optimal score.

Bock and Rautenbach~\cite{BockRautenbach19} studied \BDR with respect to \match and \vc.
They presented a characterization for \BDR with respect to \match: 
a bi-sequence has a bipartite realization with a matching of a certain size if and only if a cubic number of inequalities are satisfied.
This characterization is based on a \emph{prefix lemma} that states 
that if there exists a realization $G$ with a matching of size $\nu$,
then there exists a realization $G'$ such that the matching is between 
the vertices that correspond to the two degree prefixes
$a_1, \ldots, a_\nu$ and $b_\nu, \ldots, b_1$, respectively
(recall that $a$ and $b$ are non-increasing).
The same characterization applies to \BDR with respect to \vc due to 
K\"{o}nig's Theorem~\cite{Konig31}. In this case, 
the \emph{prefix lemma} states that if there exists a realization $G$ with a vertex
cover of size $\nu$, then there exists a realization $G'$ such that the vertex 
cover is composed of the vertices that correspond to the degree prefixes
$a_1,\ldots,a_k$ and $b_1,\ldots,b_{\nu-k}$, where $k \in \set{0,\ldots,\nu}$.
Bock and Rautenbach~\cite{BockRautenbach19} also gave a realization algorithm 
with respect to \match and \vc that relies on a \textsc{Maximum Flow} algorithm.
It should be noted that an optimal realization with respect to \vc is also optimal with respect to \is.
Hence, the above results also apply to \is.
Moreover, it is not hard to show that an optimal realization 
of $(a',b')$ with respect to \is, where $a'_i = m -a_i$, for every $i$ and 
$b'_j = n - b_j$, for every $j$, implies an optimal realization of $(a,b)$
with respect to \biclique. 
Therefore, a prefix lemma holds for this problem as well.

\subsubsection{Our Results.}
In this paper, we study the problem of \BDRlong with respect to 
\domsetlong (\domset),
where the goal is to compute a realization of the input 
bi-sequence $(a,b)$ with the smallest possible dominating set among all the 
realizations of $(a,b)$.

Our first result (\Cref{sec:realization}) is an $O(n^3m^3(n+m))$-time realization 
algorithm for \BDR with respect to \domset. 
A natural approach to solving the problem would have been to use a prefix lemma 
to reduce the search space as was done in the case of \BDR with respect to 
\match and \vc~\cite{BockRautenbach19}.
However, as we show, the prefix lemma does not hold in the case of \domset.
Instead, we utilize a less strict lemma, termed the \emph{interval lemma}. 
This lemma states that one may assume that the minimum dominating set is composed 
of two intervals, one from each partite set.
Note that given $\gamma$, there are $O(nm \gamma)$ candidate interval pairs.
Given a candidate dominating set consisting of two intervals, our algorithm 
constructs a flow network and uses a \textsc{Maximum Flow} algorithm to obtain
a bipartite realization $G$ where the candidate two intervals are a dominating 
set of $G$ or to decide that such a realization does not exist. 
We note that our flow network is not as simple as the one that is used to find an arbitrary realization for a bi-sequence or to solve \BDR with respect to \vc. 
We use a construction (i.e., a gadget) 
that ensures that the candidate two intervals are indeed a dominating set.

Our second contribution (\Cref{sec:characterization}) is a characterization for 
the existence of a realization with a dominating set of a given size $\gamma$ consisting 
of $O(nm(n+m) \gamma)$ inequalities.
This result also relies on the above mentioned flow network.
We show that a realization exists if and only if the maximum flow value in the network 
is equal to $\sum_{i=1}^n a_i = \sum_{j=1}^m b_j$. Due to the Max-flow Min-cut
Theorem, this is equivalent to requiring that the capacity of all the cuts is 
at least $\sum_{i=1}^n a_i$. We show that given two candidate intervals, this set of an exponential number of cut constraints is in fact equivalent to a system of $O(n+m)$ inequalities. Using the characterization, the time complexity of the algorithm can be reduced to $O(n^2m^2\log(\min\{n,m\}))$.

The described flow technique and a characterization implied by it were already developed for the dominating set in general graphs in~\cite{BKPR26-domset}. So the main contributions of this paper are the interval lemma and adaptation of known techniques to the bipartite setting.
\subsubsection{Further Related Work.}
Optimized degree realizations were studied for general graphs.
Rao~\cite{Rao79} gave a characterization of sequences which can be realized by a (general) 
graph containing $K_\ell$ (i.e., a clique of size $\ell$). 
This result was based on a suitable prefix lemma, which for \clique says that 
if $d$ has a realization containing $K_\ell$, then it has a realization such that the 
$\ell$ vertices of maximum degree induce a clique $K_\ell$. 
This also yields a polynomial-time realization algorithm for \DR with respect to \clique (see, e.g., \cite{Yin12,GHR18}).
Recalling that given a graph $G = (V,E)$, \is is equivalent to \clique in the complement graph 
$\bar{G} = (V,\bar{E})$, where $\bar{E} = \set{(u,v) \mid (u,v) \not\in E}$, 
it follows that a realization algorithm also exists for \DR w.r.t. \is. 
Furthermore, since $V \setminus I$ is a vertex cover if $I$ is an independent set, 
it follows that the realization algorithm with respect to \is,
also works for \DR with respect to \vc.
Gould et al.~{\cite{GJL99}} extended Rao's prefix lemma to every subgraph.
That is, they showed that if $d$ has a realization containing a subgraph $H$
then there exists a realization of $d$ containing $H$ in which the vertices of $H$ 
constitute a prefix.
Note also that \vc satisfies a prefix lemma, as does \domset~\cite{GHR18}.
For \domset, the special case where $d_1 = O(1)$ was handled in~\cite{GHR18}, 
and the general problem was recently resolved in~\cite{BKPR26-domset}.
We note that the realization algorithm for general graphs~{\cite{BKPR26-domset}} is based on 
a technique by Fulkerson et al.~{\cite{FHM1965}}, which uses a reduction to a special case of bipartite graphs 
that are symmetric and thus are amenable to a prefix lemma.
\DR with respect to \match was solved in~\cite{EKMT24,BKPR26-matching}.

Optimized realization problems were also studied in forests and trees.
A characterization for \DR in forests with respect to \domset was given in~\cite{GHR18}.
Characterizations for forest realizations that minimize the maximum independent set and 
that maximize the minimum dominating set were presented in~\cite{GHR16}.

\section{Preliminaries}

Given two integers $i, j$, define $[i,j] = \set{i,\ldots,j}$ if $i\le j$ and $[i,j] = \emptyset$ if $i>j$.
Consider a simple undirected graph $G=(V,E)$.
For a vertex $v\in V$, let $N_G(v)$ denote the set of \emph{neighbors} of $v$ in $G$, and let $\deg_G(v) = |N_G(v)|$ denote the \emph{degree} of $v$ in $G$.
(We sometimes omit the subscript $G$ when it is clear from the context.)
For vertices $v,w\in V$, denote the set of neighbors of $v$ that are not adjacent to $w$ by $N_{v-w} = N(v)\setminus N(w)$.

The set $D\subseteq V$ is a \emph{dominating set} of $G$ if every vertex $v\not\in D$ has a neighbor in $D$.
Consider two non-increasing integer sequences $a = (a_1, \ldots, a_n)$ and $b = (b_1, \ldots, b_m)$ 
such that $\sum_{i=1}^n a_{i} = \sum_{j=1}^m b_j$. 
A bipartite graph $G = (V, W, E)$ realizes the sequence pair $(a, b)$ 
if $V = \{v_1, \ldots, v_n\}$, $W = \{w_1, \ldots, w_m\}$, and moreover, 
$\deg_G(v_i) = a_i$ for every $i \in [1,n]$ and $\deg_G(w_j) = b_j$ for every $j \in [1,m]$. Define an index function $\iota(v_i) = i$ and $\iota(w_j) = j$ for any $v_i \in V$ or $w_j \in W$.

\tikzset{
  roundNode/.style={
    circle, draw, minimum size=1cm, align=center, font=\Large
  },
  dots/.style={
    draw=none, fill=none, yshift=0.1cm
  },
  edge/.style={
    font=\large
  }
}

\section{\domset Realizations for Bigraphic Sequence Pairs}
\label{sec:realization}

\begin{wrapfigure}{r}{3.5cm}
    \centering
\tikzset{
  roundNode/.style={
    circle, draw, minimum size=1cm, align=center, font=\Large
  },
  highlighter/.style={
    circle, draw, dashed, minimum size=0.5cm, inner sep=-0.2cm, align=center, font=\Large
  },
  edge/.style={
    font=\large
  }
}
\resizebox{.2\textwidth}{!}{ %
\begin{tikzpicture}[node distance=2cm and 3cm, on grid, auto]

  \node[roundNode] (x1) {$x_1$};
  \node[roundNode, below=of x1] (x2) {$x_2$};
  \node[roundNode, below=of x2] (x3) {$x_3$};

  \node[roundNode, right=of x2] (y2) {$y_2$};
  \node[roundNode, right=of x3] (y3) {$y_3$};
  \node[roundNode, right=of x1] (y1) {$y_1$};

    \node[highlighter, fit=(y1)] (highlight1) {};
    \node[highlighter, fit=(x2)] (highlight2) {};
    
  \draw (x1)[very thick, dashed] -- (y1);
  \draw (x1) -- (y2);
  \draw (x1) -- (y3);
  \draw[very thick, dashed] (x2) -- (y2);
  \draw[very thick, dashed] (x2) -- (y3);
  \draw (x3)[very thick, dashed] -- (y1);

\end{tikzpicture}}
    \caption{Prefix lemma counterexample}
    \label{fig:counterexample}
\end{wrapfigure}
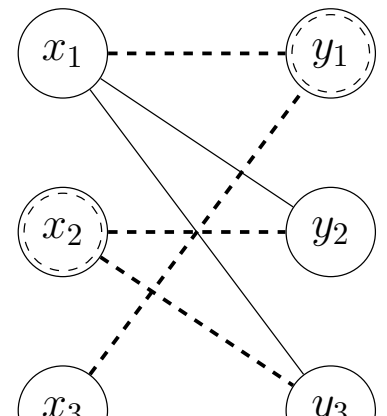
Consider two non-increasing integer sequences $a = (a_1, \ldots, a_n)$ and $b = (b_1, \ldots, b_m)$ such that $\sum_{i=1}^n a_{i} = \sum_{j=1}^m b_j$. We seek a bipartite graph $G = (V, W, E)$ with $V = \{v_1, \ldots, v_n\}$ and $W = \{w_1, \ldots, w_m\}$
that realizes the sequences $(a, b)$, 
and moreover, has the smallest possible dominating set 
for $(a,b)$. 
A \emph{configuration} is a pair $(G, D)$, where $G = (V, W, E)$ is a bipartite realization described above 
and $D$ is a (not necessarily minimum) dominating set in $G$.
A natural approach for attacking the problem of \domset realization by a bipartite graph 
is to follow the approach used for the problem of \vc realization,
i.e., establish and use a Prefix Lemma in order to narrow down the search space. 
Unfortunately, as demonstrated next, while the Prefix Lemma does hold for \domset realizations by \emph{general} graphs, it does not hold for \domset realizations by \emph{bipartite} graphs.
To see this, consider the sequences $a = (3, 2, 1)$ and $b = (2, 2, 2)$. 
The singular realization of this sequence pair yields a minimum dominating set $D = \{x_2, y_1\}$. 
(See \Cref{fig:counterexample}.)
Given that this configuration 
is the sole realization of the graph, the Prefix Lemma does not hold in this case.

\subsection{The Interval Lemma}

As the Prefix Lemma is unavailable,
we next present an alternative, hereafter referred to as the \emph{Interval Lemma}. 
Although it is less effective in terms of pinpointing the candidate sets for \domset, 
it still narrows the search space sufficiently to lead to a polynomial-time algorithm for identifying an \domset realization.

Consider a configuration $(G,D)$ and let $D_A = D \cap V$ and $D_B = D \cap W$ (i.e., $D = D_A \cup D_B$).
Call $D$ a \emph{contiguous} dominating set
if the vertices of $D_A$ and $D_B$ form contiguous subsequences of $V$ and $W$, respectively, namely, 
$D_A = \set{ v_\ell, \ldots, v_{\ell + \gamma_\ell - 1}}$ and 
$D_B= \set{w_r, \ldots, w_{r + \gamma_r - 1}}$.
A vertex $v \in D$ is \emph{self-dominated}
if each of its nondominating neighbors $w \in N(v)\setminus D$ is dominated not only by $v$ but also by some other vertex $u \in D$, $u \neq v$. 

We make use of a graph modification operation referred to as \emph{$v'$-$v$ neighbor swap}.
Assuming $\deg_G(v') \ge \deg_G(v)$, note that 
\(
|N_{v'-v}| \ge \deg_G(v') - \deg_G(v) + |N_{v-v'}| \ge |N_{v-v'}|
\).
Therefore, there is a subset $S \subseteq N_{v'-v}$ of size $|S|=|N_{v-v'}|$. Obtain $G'$ from $G$ by removing the $2|S|$ edges 
\[
\set{(v,p) \mid p \in N_{v-v'} \} ~\cup~ \{ (v',q) \mid q \in S }
~,
\]
and adding the $2|S|$ edges 
\[
\set{ (v',p) \mid p \in  N_{v-v'}\} \cup \{ (v,q) \mid q \in S}
~.
\]
One can verify that this modification does not change the degrees of the involved nodes. Moreover, $N_{G}(v) \subseteq N_{G'}(v')$.

We also make use of the following potential function 
on configurations $(G,D)$:
\begin{align*}
    S(G,D) ~=~ \sum_{v \in D_A} \iota(v) - \sum_{j=1}^{|D_A|} j + \sum_{w \in D_B} \iota(w) - \sum_{j=1}^{|D_B|} j~.
\end{align*}

\begin{lemma}\label{lem: better configuration}
    If there exists a non-contiguous configuration $(G,D)$, then there exists a configuration $(G', D')$, such that $|D'| \le |D|$ and $S(G',D') < S(G,D)$. 
\end{lemma}
\begin{proof}
Since $(G,D)$ is not contiguous, there exist vertices $\hat{v}, v'$ and $v$ on the same partite set of $G$, such that $\iota(\hat{v}) <\iota(v') < \iota(v)$ and $\hat{v}, v \in D$ while $v' \notin D$. If $\deg_G(v') = \deg_G(v)$ one can simply replace their neighbors and replace $v$ with $v'$ in the dominating set, so hereafter assume $\deg_G(\hat{v}) \ge \deg_G(v') > \deg_G(v)$. Without loss of generality assume that $\hat{v}, v',v \in V$, so $\hat{v}, v \in D_A$. Moreover, let $v$ be a vertex with the highest index in $D_A$, so $\iota(v) > |D_A|$.

The proof considers several cases, reducing harder scenarios to easier ones.
\begin{description}
\item[Case 1:]
$v$ is self-dominated. 
\\
Consider two subcases.
\begin{description}
\item[Subcase 1a:] Assume $v$ is adjacent to a vertex $w \in D_B$. Then we can simply construct a configuration $(G',D')$, where $G' = G$ and $D' = D \setminus \{v\}$. The potential function changes accordingly to $S(G', D')$,
\begin{align*}
    S(G',D') - S(G,D) = |D_A| - \iota(v) < 0
\end{align*}
\item[Subcase 1b:] If $v$ is adjacent only to nondominating vertices, then let $w$ be a neighbor of $v$ with the smallest index and let $w'$ be a vertex with the smallest index in $W \setminus D$. Note that $\iota(w') \le \iota(w)$ and $\iota(w') \le |D_B| + 1$.
        
First, we construct a graph $G'$ from $G$ such that $v$ is adjacent to $w'$ in $G'$. If $v$ and $w'$ are adjacent in $G$, then we can simply take $G' = G$. Otherwise, since $\deg_G(w') \ge \deg_G(w)$ and $v$ is adjacent to $w$, but not to $w'$, it follows that there exists $\tilde{v} \in V$ that is adjacent to $w'$, but not to $w$. Then construct $G'$ from $G$ by replacing edges $(v, w)$, $(\tilde{v}, w')$ with edges $(v,w')$, $(\tilde{v}, w)$.

Obtain $D'$ from $D$ by replacing $v$ with $w'$. Since $v$ is self-dominated, it is not hard to see that $(G', D')$ is a valid configuration. The potential function changes correspondingly,
\begin{align*}
S(G',D') - S(G,D) = |D_A| - \iota(v) + \iota(w') - (|D_B|+1) < 0
\end{align*}
\end{description}
    
\item[Case 2:]
$v'$ is adjacent to every nondominating vertex in $W$. 
\\
Since $v'$ is nondominating, it is adjacent to at least one dominating vertex in $W$ and therefore $\deg_G(v') \ge |W \setminus D_B| + 1$.
    
Obtain $G'$ from $G$ by performing $\hat{v}$-$v'$ neighbors swap. Next we show that $D$ remains a dominating set in $G'$. Since $v'$ is adjacent to every nondominating vertex in $W$ in $G$, it follows that $\hat{v}$ has the same property in $G'$, $W \setminus D \subseteq N_{G'}(\hat{v})$, so all the vertices in $W$ are dominated. Note that $\deg_{G'}(v') = \deg_G(v') \ge |W \setminus D_B| + 1$, so it is adjacent to some vertex in $D_B$ in $G'$. The rest of the vertices in $V$ have the same neighbors in $G'$ as in $G$ and therefore $D$ is a dominating set in $G'$. Moreover, since $\hat{v}$ dominates all the vertices in $W \setminus D$, it follows that $v$ is self-dominated, so the configuration $(G',D)$ can be modified according to Case 1 decreasing the potential function.

\item[Case 3:]
$v$ is not self-dominated and $v'$ is not adjacent to some nondominating vertex $w'$. \\
Let $\tilde{v}$ be a dominating neighbor of $w'$ and $\tilde{w}$ be a dominating neighbor of $v'$. Since $v$ is not self-dominated, there exists a vertex $w \in W\setminus D$ that is dominated only by $v$. Note that it could be that $w' = w$ and then $v = \tilde{v}$. To keep track of all the constraints see Figure \ref{fig: case 3} below. Highlighted nodes are dominating and the remaining nodes are nondominating. Furthermore, solid lines represent edges that exist in $G$ and dashed lines represent edges that do not.
We finish the proof considering three subcases.

    \begin{figure}[htbp]
        \centering
        \begin{subfigure}{0.32\textwidth}
            \centering
        \tikzset{
          roundNode/.style={
            circle, draw, minimum size=1cm, align=center, font=\Large
          },
          highlighter/.style={
            circle, draw, dashed, minimum size=0.5cm, inner sep=-0.2cm, align=center, font=\Large
          },
          edge/.style={
            font=\large
          }
        }
        \resizebox{.5\textwidth}{!}{ %
        \begin{tikzpicture}[node distance=2cm and 3cm, on grid, auto]
        
          \node[roundNode] (v') {$v'$};
          \node[roundNode, below=of v'] (v) {$v$};
        
          \node[roundNode, right=of v'] (tw) {$\tilde{w}$};
          \node[roundNode, right=of v] (w) {$w$};
        
            \node[highlighter, fit=(v)] (highlight1) {};
            \node[highlighter, fit=(tw)] (highlight2) {};
            
          \draw (v) -- (w);
          \draw (v) -- (tw);
          \draw (v') -- (tw);
        
        \end{tikzpicture}}
            \caption{Subcase 3a}
            \label{fig: subcase 3a}
        \end{subfigure}
        \hfill
        \begin{subfigure}{0.32\textwidth}
            \centering
        \tikzset{
          roundNode/.style={
            circle, draw, minimum size=1cm, align=center, font=\Large
          },
          highlighter/.style={
            circle, draw, dashed, minimum size=0.5cm, inner sep=-0.2cm, align=center, font=\Large
          },
          edge/.style={
            font=\large
          }
        }
        \resizebox{.5\textwidth}{!}{ %
        \begin{tikzpicture}[node distance=2cm and 3cm, on grid, auto]
        
          \node[roundNode] (v') {$v'$};
          \node[roundNode, below=of v'] (v) {$v$};
        
          \node[roundNode, right=of v'] (tw) {$\tilde{w}$};
          \node[roundNode, above=of tw] (w') {$w'$};
          \node[roundNode, right=of v] (w) {$w$};
        
            \node[highlighter, fit=(v)] (highlight1) {};
            \node[highlighter, fit=(tw)] (highlight2) {};
            
          \draw (v)[very thick, dashed] -- (tw);
          \draw (v')[very thick, dashed] -- (w');
          \draw (v) -- (w);
          \draw (v) -- (w');
          \draw (v') -- (tw);

        \end{tikzpicture}}
            \caption{Subcase 3b}
            \label{fig: subcase 3b}
        \end{subfigure}
        \hfill
        \begin{subfigure}{0.32\textwidth}
            \centering
        \tikzset{
          roundNode/.style={
            circle, draw, minimum size=1cm, align=center, font=\Large
          },
          highlighter/.style={
            circle, draw, dashed, minimum size=0.5cm, inner sep=-0.2cm, align=center, font=\Large
          },
          edge/.style={
            font=\large
          }
        }
        \resizebox{.5\textwidth}{!}{ %
        \begin{tikzpicture}[node distance=2cm and 3cm, on grid, auto]
        
          \node[roundNode] (tv) {$\tilde{v}$};
          \node[roundNode, below=of tv] (v') {$v'$};
          \node[roundNode, below=of v'] (v) {$v$};

          \node[roundNode, right=of tv] (w') {$w'$};
          \node[roundNode, right=of v'] (tw) {$\tilde{w}$};
          \node[roundNode, right=of v] (w) {$w$};
        
            \node[highlighter, fit=(v)] (highlight1) {};
            \node[highlighter, fit=(tw)] (highlight2) {};
            \node[highlighter, fit=(tv)] (highlight3) {};
            
          \draw (v)[very thick, dashed] -- (tw);
          \draw (v')[very thick, dashed] -- (w');
          \draw (v)[very thick, dashed] -- (w');
          \draw (tv)[very thick, dashed] -- (w);
          \draw (v) -- (w);
          \draw (v') -- (tw);
          \draw (tv) -- (w');
        
        \end{tikzpicture}}
            \caption{Subcase 3c}
            \label{fig: subcase 3с}
        \end{subfigure}
        \caption{Illustration to Case 3}
        \label{fig: case 3}
    \end{figure}
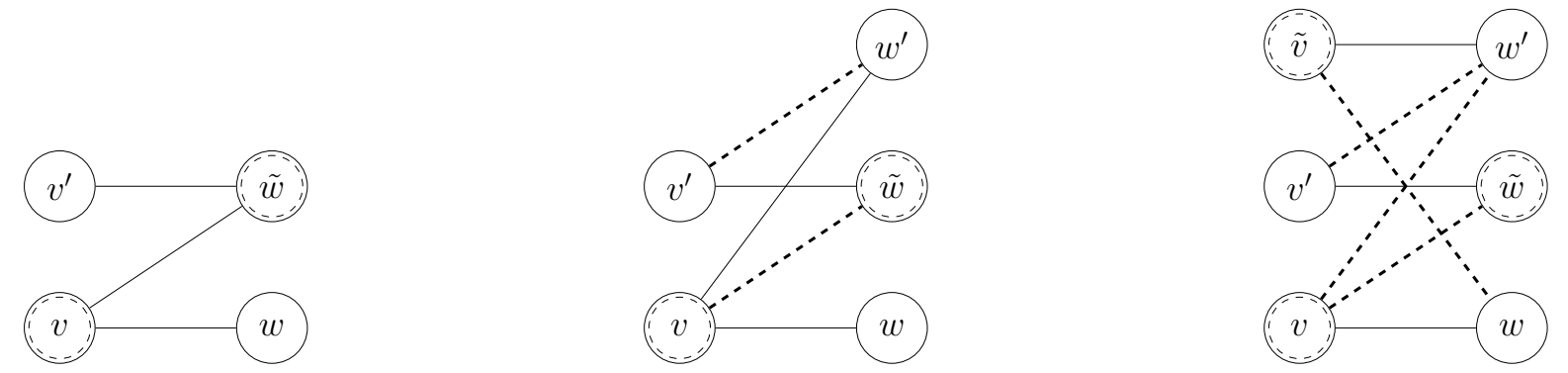

\begin{description}
\item[Subcase 3a:] Assume $v$ is adjacent to $\tilde{w}$. Then obtain $G'$ by applying $v'$-$v$ neighbors swap and let $D' = (D \cup \{v'\} )\setminus \{v\}$. Note that $D'$ is a dominating set in $G'$. Indeed, all the vertices in $N_G(v)$ are dominated by $v'$ in $G'$ and $v$ itself is dominated by $\tilde{w}$. The potential function changes accordingly to $S(G', D')$,
\begin{align*}
S(G', D') - S(G,D) = \iota(v') - \iota(v) < 0~.
\end{align*}
\item[Subcase 3b:] Assume $v$ is adjacent to $w'$, but not to $\tilde{w}$. Again obtain $G'$ by applying $v'$-$v$ neighbors swap, but make sure this time to swap $\tilde{w}$ (i.e., choose $S$ containing $\tilde{w}$). Note that $N_{v-v'}$ is non-empty because $w'$ is adjacent to $v$, but not to $v'$, so $S$ is non-empty too. The rest of the proof follows Subcase 3a. Then $D'$ is constructed from $D$ by swapping $v$ with $v'$ as before. We end up with a configuration $(G', D')$ and the value of the potential function $S(G', D')$ decreases as in Subcase 3a.  
\item[Subcase 3c:] Assume $v$ is neither adjacent to $w'$, nor to $\tilde{w}$. Construct $G'$ from $G$ by replacing edges $(\tilde{v}, w')$, $(v, w)$ with edges $(\tilde{v}, w)$, $(v, w')$. It is not hard to verify that $(G', D)$ is a valid configuration. Since $v$ is adjacent to $w'$ in $G'$, we can proceed according to Subcase 3b. 
\end{description}
\end{description}
\qed\end{proof}

Finally, we state and prove our interval lemma.

\begin{lemma}[\bf{Interval Lemma}] 
\label{lem: Interval lemma}
If there exists a configuration $(G,D)$, then there also exists a configuration $(G', D')$ where $D'$ is a contiguous dominating set of size at most $|D|$.
\end{lemma}

\begin{proof}
Note that the potential function $S(G,D)$ always returns an integral value and is bounded from above and from below by
\begin{align*}
-\sum_{i=1}^n i - \sum_{j=1}^m j ~\le~ S(G,D) 
~\le~ \sum_{i=1}^n i + \sum_{j=1}^m j~.
\end{align*}
Take a configuration $(G',D')$ with the minimum value of $S(G',D')$ 
and $|D'| \le |D|$. By \Cref{lem: better configuration} $D'$ must be contiguous.
\qed\end{proof}

\subsection{Reduction to Flow}

To find the desired realization minimizing the dominating set, 
we go through every possible assignment of values to 
$(\ell, \gamma_\ell, r,\gamma_r)$
in the Interval Lemma \ref{lem: Interval lemma}, 
and check if a suitable realizing bipartite graph $G =(V,W,E)$ exists with 
$V = \{v_1, \ldots, v_n\}$ and $W = \{w_1, \ldots, w_m\}$ and dominating set 
$D = \set{v_\ell,\ldots,v_{\ell+\gamma_\ell - 1}} \cup \set{w_r,\ldots,w_{r + \gamma_r - 1 }}$,
by reducing the problem to a flow problem. 
Specifically, given $(a, b)$ and a quadruple $Q = (\ell, \gamma_\ell, r, \gamma_r)$, 
we construct a directed bipartite flow graph $G_{a, b,Q} =(\tV,\tE)$ as follows.

The vertex set $\tV$ of $G_{a, b,Q}$ includes a \emph{source} $s$, 
a \emph{sink} $t$ and two sets $X = \{x_1,\ldots, x_n\}$ and $Y = \{y_1,\ldots, y_m\}$ 
that correspond to the vertices of $V$ and $W$, respectively.
These sets are further partitioned into two sets each, according to their intended role:
the sets 
\begin{align*}
X_D & =\{x_i \mid i \in [\ell,\ell + \gamma_\ell - 1]\}
    & \text{and} 
    & &
Y_D &= \{y_j \mid j \in [r,r + \gamma_r - 1]\}
\end{align*}
contain the candidates to be the dominating vertices, 
and the sets $X_S = X \setminus X_D$ and $Y_S = Y \setminus Y_D$ 
contain the remaining (passive) vertices, which are to be dominated. 
In addition, $\tV$ contains two additional sets of supplementary vertices, 
$X'_S = \{x'_i \mid x_i \in X_S\}$ and $Y'_S = \{y'_j \mid y_j \in Y_S\}$, 
used to enforce domination constraints.

Overall, the vertex set is
\[ \tV ~=~ \{s, t\} ~\cup ~ X ~\cup~ Y ~\cup~ X'_S ~\cup~ Y'_S \]
The edges are capacitated and can be conveniently thought of as directed from left to right, i.e., an edge $(\alpha, \beta, \gamma)$ leads from the node $\alpha$ to the node $\beta$ and can carry up to $\gamma$ units of flow. The source $s$ has edges leading to every node $x_i \in X$, with capacity $a_i$, to enforce the degree constraint for the corresponding vertex $v_i$ in the realizing graph $G$. Similarly, there are edges leading from every node $y_j \in Y$ to the sink $t$, with capacity $b_j$, to enforce the degree constraint for the vertices $w_j$ in $G$. 

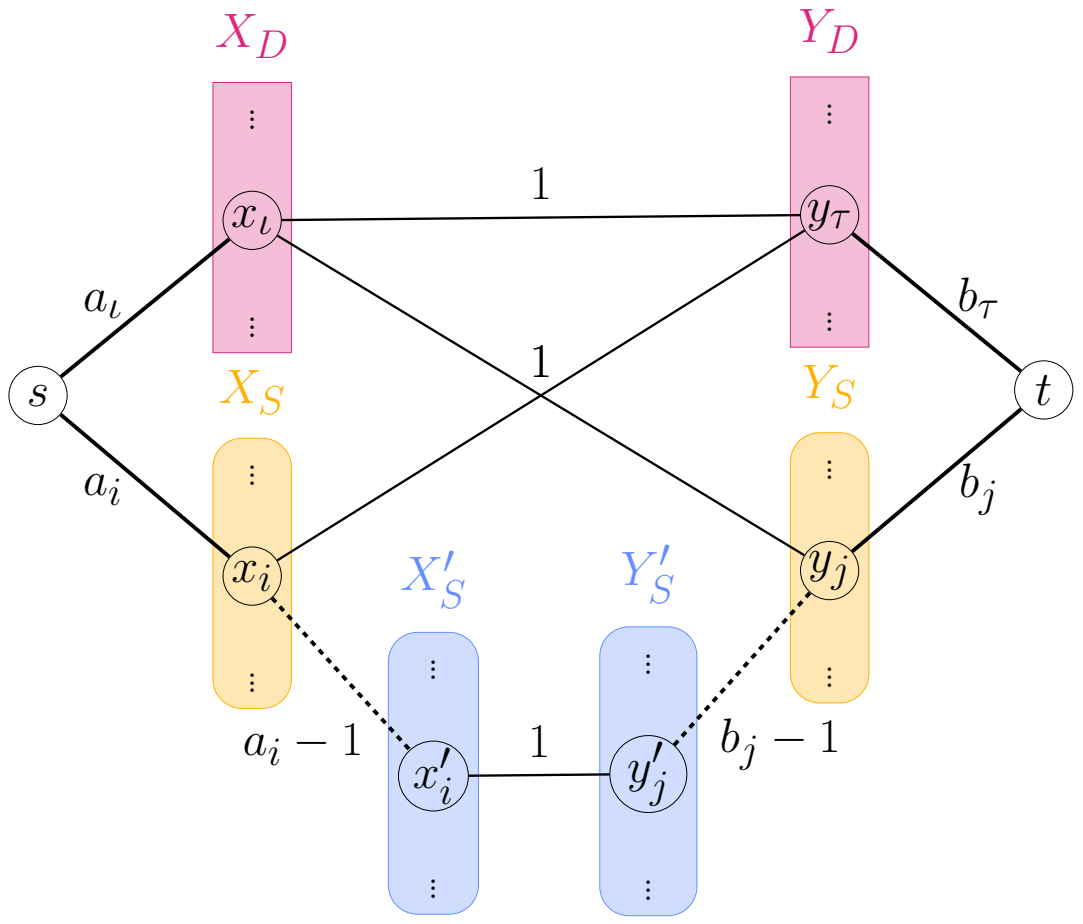
\begin{figure}[t]%
\centering
\resizebox{0.6\textwidth}{!}{
\tikzset{
myptr/.style={-{Stealth[scale=1.5]}},
}

\definecolor{r}{HTML}{DC267F}
\definecolor{g}{HTML}{FFB000}
\definecolor{b}{HTML}{648FFF}
\def\thicc{1.2}
\def\Thicc{2.0}

\begin{tikzpicture}[
    roundnode/.style={circle, draw=black, minimum size=30pt, inner sep=0pt},
    squarebox/.style={draw=r, fill=r, fill opacity=0.3, inner sep=5pt},
    roundedbox/.style={draw=g, fill=g, fill opacity=0.3, rounded corners=15pt, inner sep=5pt}, 
    roundedboxprime/.style={draw=b, fill=b, fill opacity=0.3, rounded corners=15pt, inner sep=5pt}, %
    font=\Huge
]

\node[dots] (x1) {$\vdots$};
\node[roundnode, below=1cm of x1] (xi) {$x_{\iota}$};
\node[dots, below=1cm of xi] (xl) {$\vdots$};

\node[dots, below=2cm of xl] (xl1) {$\vdots$};
\node[roundnode, below=1cm of xl1] (xi') {$x_i$};
\node[dots, below=1cm of xi'] (xn) {$\vdots$};

\node[color=r, above=0.5cm of x1] (XD) {$X_D$};
\node[color=g, above=0.5cm of xl1] (XS) {$X_S$};

\node[fit=(x1)(xn)] (xcenter) {};

\node[roundnode, left=3cm of xcenter] (s) {$s$};

\node[dots, below right=4cm of xl1] (x'1) {$\vdots$};
\node[roundnode, below=1cm of x'1] (x'i) {$x'_i$};
\node[dots, below=1cm of x'i] (x'n) {$\vdots$};

\node[color=b, above=0.5cm of x'1] (X'S) {$X'_S$};

\node[dots, right=10cm of x1] (y1) {$\vdots$};
\node[roundnode, below=1cm of y1] (yj) {$y_{\tau}$};
\node[dots, below=1cm of yj] (yk) {$\vdots$};

\node[dots, below=2cm of yk] (yk1) {$\vdots$};
\node[roundnode, below=1cm of yk1] (yj') {$y_j$};
\node[dots, below=1cm of yj'] (ym) {$\vdots$};

\node[color=r, above=0.5cm of y1] (YD) {$Y_D$};
\node[color=g, above=0.5cm of yk1] (YS) {$Y_S$};

\node[fit=(y1)(ym)] (ycenter) {};

\node[roundnode, right=3cm of ycenter] (t) {$t$};

\node[dots, below left=4cm of yk1] (y'1) {$\vdots$};
\node[roundnode, below=1cm of y'1] (y'j) {$y'_j$};
\node[dots, below=1cm of y'j] (y'm) {$\vdots$};

\node[color=b, above=0.5cm of y'1] (Y'S) {$Y'_S$};

\draw[line width=\Thicc] (s) -- (xi) node[pos=0.25, above=0.3cm] {$a_{\iota}$};
\draw[line width=\Thicc] (s) -- (xi') node[pos=0.25, below=0.3cm] {$a_i$};

\draw[line width=\Thicc] (yj) -- (t) node[pos=0.75, above=0.2cm] {$b_{\tau}$};
\draw[line width=\Thicc] (yj') -- (t) node[pos=0.75, below=0.2cm] {$b_j$};

\draw[line width=\thicc] (xi) -- (yj) node[pos=0.5, above=0.2cm] {$1$};
\draw[line width=\thicc] (xi) -- (yj') node[pos=0.5, above=0.2cm] {$1$};
\draw[line width=\thicc] (xi') -- (yj) node[pos=0.5, above=0.2cm] {$1$};

\draw[line width=\thicc] (x'i) -- (y'j) node[pos=0.5, above=0.2cm] {$1$};
\draw[line width=\Thicc, dashed] (xi') -- (x'i) node[pos=0.75, below left=0.1cm] {$a_i-1$};
\draw[line width=\Thicc, dashed] (y'j) -- (yj') node[pos=0.25, below right=0.1cm] {$b_j-1$};

\begin{scope}[on background layer]
    \node[squarebox, fit=(x1) (xi) (xl)] {};
    \node[roundedbox, fit=(xl1) (xi') (xn)] {};
    \node[roundedboxprime, fit=(x'1) (x'i) (x'n)] {};
    \node[squarebox, fit=(y1) (yj) (yk)] {};
    \node[roundedbox, fit=(yk1) (yj') (ym)] {};
    \node[roundedboxprime, fit=(y'1) (y'j) (y'm)] {};
\end{scope}

\end{tikzpicture}}
\caption{The flow network
}
\label{fig:flow graph}
\end{figure}

Additional edges are employed within gadgets designed to ensure the correctness of the reduction and the domination requirements. To describe the construction in a concise manner, we write $\hat{X} \hourglass \hat{Y}$ if the vertex sets $\hat{X}$ and $\hat{Y}$ induce a complete bipartite subgraph. The different vertex sets are connected by four such bipartite subgraphs: $X_D \hourglass Y_D$, $X_D \hourglass Y_S$, $X_S \hourglass Y_D$ and $X'_S \hourglass Y'_S$, with each edge having capacity 1. Finally, every $x_i \in X_S$ is connected to the corresponding $x'_i \in X'_S$ with capacity $a_i - 1$. In the same way, every $y'_j \in Y'_S$ is connected to corresponding $y_j \in Y_S$ by an edge of capacity $b_j - 1$. Hereafter, denote these connections by $X_S \equiv X'_S$ and $Y'_S \equiv Y_S$. Overall, the connections are depicted schematically in Figure
\ref{fig:flow graph}. We maintain the following notation: edges with capacity $1$ are thin and edges in subgraphs $X_S \equiv X'_S$ and $Y'_S \equiv Y_S$ are dashed.

Next, we calculate the maximum flow from the source $s$ to the sink $t$ in the flow graph $G_{a,b,Q}$.
The following lemma establishes the connection between the flow graph and the graph $G$ realizing $(a, b)$.

\begin{lemma} 
There exists a bipartite graph $G = (V, W, E)$, 
where $V = \{v_1, \ldots, v_n\}$ and $W = \{w_1, \ldots, w_m\}$, 
that realizes the degree sequence pair $(a, b)$, 
such that $D = D_A \cup D_B$ with $D_A = \{v_\ell, \ldots, v_{\ell+\gamma_\ell-1}\}$
and $D_B =  \{w_r, \ldots, w_{r +\gamma_r - 1}\}$ 
forms a dominating set for $G$ if and only if the 
maximum $s$-$t$ flow value in $G_{a, b,Q}$ is equal to $\sum_{i=1}^n a_i$.
\end{lemma}

\begin{proof}
Suppose the maximum $s$-$t$ flow value in $G_{a, b,Q}$ is $\sum_{i=1}^n a_i$. 
By the integral flow theorem we can assume that the maximum flow is integral.
Define the bipartite graph $G = (V, W, E)$ over $V = \{v_1, \ldots, v_n\}$ 
and $W = \{w_1, \ldots, w_m\}$ by selecting the edge set $E$ as follows.
For every $i \in [1,n]$ and $j \in [1,m]$, 
if there is some edge $e\in\tE$ of the form $(x_i, y_j)$ or $(x'_i, y'_j)$ with $\FLOW(e)=1$, 
then add the edge $(v_i, w_j)$ to $E$.

To verify that $G$ correctly realizes $(a, b)$, we first observe that the edges selected to $E$ form a graph (rather than a multigraph), because for any pair $i \in [1,n]$ and $j \in [1,m]$, 
$\tE$ contains either only an edge $(x'_i, y'_j)$ (if $x_i \in X_S$ and $y_j \in Y_S$) or only $(x_i, y_j)$ (otherwise), but not both. 

Next, we show that $\deg_G(v_i)=a_i$ for every $i \in [1,n]$ and $\deg_G(w_j)=b_j$ for every $j \in [1,m]$. 
Note that the total flow from $s$ to $t$ is $\sum_{i=1}^n a_i = \sum_{j=1}^m b_j$, 
and therefore all edges from $s$ to the vertices of $X$ and from the vertices of $Y$ to $t$ are saturated.
We consider four cases:
\begin{itemize}
\item For $v_i \in D_A$, the degree of $v_i$ in $G$, $\deg_G(v_i)$, 
      is determined by the number of edges $(x_i, y_j)$ with $\FLOW(x_i, y_j)=1$ for $j \in [1,m]$. 
      Since $x_i$ receives a flow of $a_i$ and outputs it in unit flows through edges $(x_i, y_j)$, 
      its degree is $a_i$.
      
\item For $w_j \in D_B$, a dual argument shows that 
$\deg_G(w_j)=b_j$.

\item For $v_i \in V \setminus D_A$, 
$\deg_G(v_i)$ is determined by the number of edges $(x_i, y_j)$ and $(x'_i, y'_j)$ with unit flow for $j \in [1,m]$.
Divide the flow emanating from $x_i$ into two parts, 
$\FLOW(s, x_i) = \phi_1+\phi_2$ where $\phi_1 = \sum_{w_j \in D_B} \FLOW(x_i, y_j)$ and $\phi_2 = \FLOW(x_i, x'_i)$. 
The first term, $\phi_1$, consists of unit flows on separate edges $(x_i, y_j)$, 
each of which contributes an edge to $v_i$, hence overall it contributes $\phi_1$ to $\deg_G(v_i)$.
The rest of the flow goes to $x'_i$. Since $x'_i$ forwards its incoming flow through edges $(x'_i, y'_j)$ with unit capacity, each of which contributes an edge to $v_i$, it increases degree of $v_i$ by $\phi_2$. Hence, the degree of $v_i$ equals $\phi_1+\phi_2= \FLOW(s, x_i) = a_i$.

\item For $w_j \in W \setminus D_B$, a dual argument shows that $\deg_G(w_j)=b_j$.
\end{itemize}

Therefore, the bipartite graph $G$ correctly realizes $(a, b)$.

It remains to show that $D$ is a dominating set. 
Since $y_j \in Y_S$ conserves flow and $\FLOW(y'_j, y_j) \leq b_j - 1$, 
there is at least one $x_i \in X_D$ with $\FLOW(x_i, y_j) = 1$. 
Similarly, for every $x_i \in X_S$ there exists at least one $y_j \in Y_D$ such that $\FLOW(x_i, y_j) = 1$. 
It follows that at least one vertex from $D$ is adjacent to each 
$v_i \in V \setminus D$
and $w_j \in W \setminus D$.
Thus, $D$ is indeed dominating set for $G$.

Conversely, suppose $G = (V, W, E)$, where $V = \{v_1, \ldots, v_n\}$ and $W = \{w_1, \ldots, w_m\}$, 
is a bipartite graph realizing $(a, b)$ such that $D$ is a dominating set for it. 
We define a flow function on $G_{a, b,Q}$ and prove that the maximum $s$-$t$ flow value equals $\sum_{i=1}^n a_i$. 
Obviously, the flow can not exceed $\sum_{i=1}^n a_i$, 
so it is enough to prove that $\sum_{i=1}^n a_i$ is achievable.
The flow is defined as follows:
\begin{enumerate}
\item \textbf{Initialization}: Set all flows to $0$.

\item \textbf{Saturate Edges from Source and to Sink}:
\\
    $\FLOW(s, x_i) \leftarrow a_i \quad \text{for every } i \in [1,n]$,
\\
    $\FLOW(y_j, t) \leftarrow b_j \quad \text{for every } j \in [1,m]$.
    
\item \textbf{Flow to Supplementary $X$-Vertices}:
\\
For every $i \in [1,n]$, such that $v_i \notin D$:
\\ - Set $N_i \gets$ number of edges $(v_i, w_j) \in E$ with $w_j \notin D$.
\\
- Set $\FLOW(x_i, x'_i) \leftarrow N_i$.
\\
Since $v_i$ is adjacent to at least one dominating vertex, $N_i \leq a_i - 1$, 
so the flows satisfy the capacity constraints.

\item \textbf{Flow from Supplementary $Y$-Vertices}:
\\
For every $j \in [1,m]$, such that $w_j \notin D$:
\\ - Set $M_j \gets$ number of edges $(v_i, w_j) \in E$ with $v_i \notin D$.
\\ - Set $\FLOW(y'_j, y_j) \leftarrow M_j$.
\\
Since $w_j$ is adjacent to at least one dominating vertex, $M_j \leq b_j - 1$, so the flows satisfy the capacity constraints.
\item \textbf{Flow Through Edges}:
\\
For every $i \in [1,n]$ and $j \in [1,m]$, such that $(v_i, w_j) \in E$:
    \begin{enumerate}
    \item If $x_i \in X_S$ and $y_j \in Y_S$, then set $\FLOW(x'_i, y'_j) \leftarrow 1$,
    \item If either $x_i \in X_D$ or $y_j \in Y_D$, then set $\FLOW(x_i, y_j) \leftarrow 1$.
    \end{enumerate}
\end{enumerate}

\subsubsection{Proof of Flow Validity.}
Note that by construction, the total flow out of the source $s$ is $\sum_{i=1}^n a_i$ and the total flow into the sink $t$ is 
$\sum_{j=1}^m b_j$. By the assumption on $a$ and $b$, these flows are equal. It remains to verify that for every vertex except $s$ and $t$, the incoming and outgoing flows are equal.

\noindent
\textbf{Node $x_i$}: 
By construction, the incoming flow at $x_i$ is $\FLOW(s, x_i) = a_i$. 
As for the outgoing flow, recall that for every $i$, there are exactly $a_i$ indices $j$ for which $(v_i,w_j)\in E$. 
If $v_i \in D_A$, then by construction the flows emanating from $x_i$ are $\FLOW(x_i, y_j)=1$ for precisely those indices, so $\sum_{j=1}^m \FLOW(x_i, y_j)=a_i$. If $v_i \notin D_A$, then the outgoing flows from $x_i$ are $\FLOW(x_i, y_j)=1$ for $w_j\in D_B$, $(v_i,w_j)\in E$ and 
\(
\FLOW(x_i, x'_i)= N_i = |\{ w_j \notin D_B \mid (v_i, w_j) \in E\}|
\).
So 
\(
\sum_{w_j \in D_B} \FLOW(x_i, y_j) + \FLOW(x_i, x'_i) = a_i
\).
Hence $x_i$ conserves flow.

\noindent
\textbf{Node $y_j$}:
A dual argument shows that
$y_j$ conserves flow.

\noindent
\textbf{Node $x'_i$}:
The incoming flow at $x'_i$ is $N_i$ by construction. The outgoing flow is 
\[
\sum_{w_j \notin D_B} \FLOW(x'_i, y'_j) = N_i
\]
by definition. Hence, $x'_i$ conserves flow.

\noindent
\textbf{Node $y'_j$}:
A dual argument shows that $y'_j$ conserves flow.

Therefore, the flow function defined above is valid, 
and the maximum $s$-$t$ flow value in $G_{a, b,Q}$ equals $\sum_{i=1}^n a_i$, 
implying the existence of a bipartite graph $G$ with $D$ as a dominating set.
\qed\end{proof}

There are $O(n^2)$ possible intervals in $V$ and $O(m^2)$ possible intervals in $W$, 
yielding $O(n^2m^2)$ candidate pairs. For each pair, the corresponding flow network has $O(n+m)$ vertices and $O(nm)$ 
edges. Using Orlin’s maximum flow algorithm \cite{Orlin13}, which runs in $O(|V||E|)$ time, we can check each candidate pair in $O(nm(n+m))$ time. Thus, the overall running time of the algorithm is $O(n^3m^3(n+m))$.
\section{Characterization}
\label{sec:characterization}
In this section, we complement the previous result by providing a Gale-Ryser type characterization for the realizability of bi-sequences with a dominating set of size $\gamma$. 

It was shown in the previous section 
that a bi-sequence $(a,b)$ has a realization 
with a dominating set of size $\gamma$ if and only if there exists 
$Q = (\ell,\gamma_\ell,r,\gamma_r)$, where $\gamma = \gamma_\ell + \gamma_r$, 
such that the flow graph $G_{a, b, Q}$ admits a flow of size 
$\sum_{i=1}^n a_i$.
Therefore, by the max--flow min--cut theorem, $(a,b)$ has a realization with a dominating set of size $\gamma$ 
if and only if the capacity of every cut in the graph $G_{a,b,Q}$ is at least 
$\sum_{i=1}^n a_i = \sum_{j=1}^m b_j$.
This can serve as a basis for the sought characterization. 
Unfortunately, a naive implementation yields a characterization of length exponential in the input size. To overcome that difficulty, we apply the following strategy.
As a first step, we show that a minimum cut belongs to one of four families of cuts 
$\calF_1$, $\calF_2$, $\calF_3$ and $\calF_4$.
Each cut in the above families induces a constraint whose size is at least 
$\sum_{i=1}^n a_i = \sum_{j=1}^m b_j$. 
Next, we decrease the number of constraints by showing that it is sufficient to 
consider $O(n+m)$ constraints per family. More specifically, we prove the following theorem.

\begin{theorem}
\label{thm:systems}
Let $(a,b)$ be a bi-sequence, where $a = (a_1, \ldots, a_n)$ and $b = (b_1, \ldots, b_m)$ 
are non-increasing sequences, and $\sum_{i=1}^n a_i = \sum_{j=1}^m b_j$.
There exists a realization with a dominating set of size $\gamma$ if and only if 
there exists two contiguous sets $\IXD \subseteq [1, n]$ and $\IYD \subseteq [1,m]$, 
where $|\IXD| + |\IYD| = \gamma$, $\IXS = [1,n] \setminus \IXD$ 
and $\IYS = [1,m] \setminus \IYD$,
such that the following systems of constraints are satisfied:
\begin{align}
\label{eqn:erdos-gallai 1}
\sum_{i \in I}(a_i-1) 
&\leq \sum_{j\in \IYD} b_j - |\IXS| + \sum_{j\in \IYS} \min\{|I|, b_j-1\}, 
~\forall~
\mbox{\rm prefix } I \subseteq \IXS \\
\label{eqn:erdos-gallai 2}
\sum_{i\in I} a_i 
& \leq \sum_{j=1}^m \min\{|I|, b_j\},
~\forall~
\mbox{\rm prefix }
I \subseteq [1,n] \\
\label{eqn:erdos-gallai 3}
\sum_{i \in I} a_i 
& \leq \sum_{j\in \IYD} \min\{|I|, b_j\} + \sum_{j \in \IYS} \min\{|I|, b_j-1\},
~\forall~
\mbox{\rm prefix } I \subseteq \IXS \\
\label{eqn:erdos-gallai 4}
\sum_{j \in J} b_j 
& \leq \sum_{i \in \IXD} \min\{|J|, a_i\} + \sum_{i \in \IXS} \min\{|J|, a_i-1\}, 
~\forall~
\mbox{\rm prefix } J \subseteq \IYS
\end{align}
\end{theorem}

Notice that System~\eqref{eqn:erdos-gallai 2} coincides with the Gale-Ryser conditions.
There are $O(nm\gamma)$ possible pairs $I_D$ and $J_D$, resulting in $O(n m (n+m) \gamma)$ inequalities for the whole characterization. 
The rest of this section is devoted to proving the above theorem.

Let $c(S,T)$ be the capacity of an $s$-$t$ cut $(S,T)$ of $G_{a,b,Q}$.
Let $\calS$ be the set of $s$-$t$ cuts of the graph $G_{a, b, Q}$.
Clearly, $\calS = \calS_{00} \cup \calS_{01} \cup \calS_{10} \cup \calS_{11}$, where
\begin{align*}
\calS_{00} & = \set{(S,T) \mid S \cap X_D = \emptyset, T \cap Y_D = \emptyset}, \\
\calS_{01} & = \set{(S,T) \mid S \cap X_D = \emptyset, T \cap Y_D \neq \emptyset}, \\
\calS_{10} & = \set{(S,T) \mid S \cap X_D \neq \emptyset, T \cap Y_D = \emptyset}, \\
\calS_{11} & = \set{(S,T) \mid S \cap X_D \neq \emptyset, T \cap Y_D \neq \emptyset}. 
\end{align*}

Our next goal is to consider smaller cut families. Let $\IXD = [\ell, \ell+\gamma_\ell-1]$ and $\IYD = [r, r + \gamma_r - 1]$ 
be indices of the vertices in $X_D$ and $Y_D$, respectively. 
Analogously, let $\IXS = [1,n] \setminus \IXD$ and $\IYS = [1,m]\setminus \IYD$ 
be the vertices of $X_S$ and $Y_S$. Define the following cut families:
\begin{itemize}
\item $\calF_1 = \set{(S^1_{I,J},V\setminus S^1_{I,J}) \mid I \subseteq \IXS, J \subseteq \IYS}$, 

      where $S^1_{I,J} = \{x_i' \mid i\in I\} \cup \{y'_j \mid j \in J\} \cup X_S \cup Y_D$.
\item $\calF_2 = \{(S^2_{I,J}, V\setminus S^2_{I,J}) \mid I \subseteq [1,n], J \subseteq[1,m]\}$, 

      where $S^2_{I,J} = \{x_i \mid i\in I\} \cup \{y_j \mid j \in J\} \cup
             \{x_i' \mid i \in I \cap \IXS \} \cup \{y'_j \mid j \in J \cap \IYS\}$.

\item $\calF_3 = \{(S^3_{I,J}, V\setminus S^3_{I,J}) \mid I \subseteq \IXS, ~ J \subseteq [1,m]\}$,

    where $S^3_{I,J} = \{x_i, ~x'_i \mid i \in I\} \cup \{y_j \mid j \in J \cap \IYD\} 
    \cup \{y'_j \mid j \in J \cap \IYS \}$.

\item $\calF_4 = \{(S^4_{I,J}, V\setminus S^4_{I,J}) \mid I \subseteq [1,n], ~ J \subseteq \IYS \}$,

    where $S^4_{I,J} = \{y_j, ~y'_j \mid j \in J\} \cup \{x_i \mid i \in I \cap \IXD\} 
    \cup \{x'_i \mid i \in I \cap \IXS \} \cup X_S \cup Y_D$.
            
\end{itemize}
The following lemmas complete the proof of \Cref{thm:systems}. Using this characterization, the running time can be improved to $O(n^2m^2 \log(\min(n,m)))$. Details of the proof can be found in the Appendix.
\begin{lemma}
\label{cor:one}
The following inequalities are satisfied:
\begin{itemize}
\item $\min \set{c(S,T) \mid (S,T) \in \calF_1} \leq \min \set{c(S,T) \mid (S,T) \in \calS_{00}}$.
\item $\min \set{c(S,T) \mid (S,T) \in \calF_2} \leq \min \set{c(S,T) \mid (S,T) \in \calS_{11}}$.
\item $\min \set{c(S,T) \mid (S,T) \in \calF_3} \leq \min \set{c(S,T) \mid (S,T) \in \calS_{01}}$.
\item $\min \set{c(S,T) \mid (S,T) \in \calF_4} \leq \min \set{c(S,T) \mid (S,T) \in \calS_{10}}$.
\end{itemize}
\end{lemma}

\begin{lemma} \label{lemma:F1-F4}
For every $i\in\{1,2,3,4\}$, the capacity of every cut in $\calF_i$ is at least $\sum_{j=1}^n a_j$ if and only if System $(i)$ in Theorem 1 is satisfied.
\end{lemma}

\def\APPENDA{
\begin{figure}[htp!]
    \centering
    \begin{subfigure}{0.45\textwidth}
    \centering
    \resizebox{1\textwidth}{!}{
    \tikzset{
    myptr/.style={-{Stealth[scale=1.5]}},
    }
    
    \definecolor{r}{rgb}{1, 0, 0.0}
    \definecolor{g}{rgb}{0.0, 0.5, 0.0}
    \definecolor{b}{rgb}{0.0, 0, 1}
    \definecolor{rg}{rgb}{1, 0.5, 0.0}
    \definecolor{gb}{rgb}{0, 0.5, 1}
    \def\thicc{1.2}
    \def\Thicc{2.}
    
    \begin{tikzpicture}[
        roundnode/.style={circle, draw=black, minimum size=35pt, inner sep=0pt, line width=1pt, font=\Huge},
        squarebox/.style={draw=black, inner sep=5pt, font=\Huge},
        roundedbox/.style={draw=black, rounded corners=15pt, inner sep=5pt, font=\Huge}, 
        roundedboxprime/.style={draw=black,rounded corners=15pt, inner sep=5pt, font=\Huge}, %
        dots/.style={draw=none, fill=none, font=\Huge},
        edgenode/.style={font=\Huge},
    ]
    
    \node[dots] (x1) {$\vdots$};
    \node[roundnode, below=1.3cm of x1] (xi) {$x_{\iota}$};
    \node[dots, below=0.7cm of xi] (xl) {$\vdots$};
    
    \node[dots, below=2cm of xl] (xl1) {$\vdots$};
    \node[roundnode, below=1cm of xl1] (xi') {$x_i$};
    \node[dots, below=1cm of xi'] (xn) {$\vdots$};
    
    \node[font = \Huge, left=0.8cm of x1] (XD) {$X_D$};
    \node[font = \Huge, above=0.5cm of xl1] (XS) {$X_S$};
    
    \node[fit=(x1)(xn)] (xcenter) {};
    \node[roundnode, left=3cm of xcenter] (s) {$s$};
    
    \node[dots, below right=4cm of xl1] (x'1) {$\vdots$};
    \node[roundnode, below=0.7cm of x'1] (x'i) {$x'_i$};
    \node[dots, below=1.3cm of x'i] (x'n) {$\vdots$};
    
    \node[font = \Huge, above=0.5cm of x'1] (X'S) {$X'_S$};
    
    \node[dots, right=10cm of x1] (y1) {$\vdots$};
    \node[roundnode, below=0.7cm of y1] (yj) {$y_{\tau}$};
    \node[dots, below=1.3cm of yj] (yk) {$\vdots$};
    
    \node[dots, below=2cm of yk] (yk1) {$\vdots$};
    \node[roundnode, below=1cm of yk1] (yj') {$y_j$};
    \node[dots, below=1cm of yj'] (ym) {$\vdots$};
    
    \node[font = \Huge, right=0.8cm of y1] (YD) {$Y_D$};
    \node[font = \Huge, above=0.5cm of yk1] (YS) {$Y_S$};
    
    \node[fit=(y1)(ym)] (ycenter) {};
    \node[roundnode, fill=black!20, right=3cm of ycenter] (t) {$t$};
    
    \node[dots, below left=4cm of yk1] (y'1) {$\vdots$};
    \node[roundnode, below=1.3cm of y'1] (y'j) {$y'_j$};
    \node[dots, below=0.7cm of y'j] (y'm) {$\vdots$};
    
    \node[font = \Huge, above=0.5cm of y'1] (Y'S) {$Y'_S$};
    
    \draw[line width=\Thicc] (s) -- (xi) node[edgenode, pos=0.25, above=0.3cm] {$a_{\iota}$};
    \draw[line width=\Thicc] (s) -- (xi') node[edgenode, pos=0.25, below=0.3cm] {$a_i$};
    
    \draw[line width=\Thicc] (yj) -- (t) node[edgenode, pos=0.75, above=0.3cm] {$b_{\tau}$};
    \draw[line width=\Thicc] (yj') -- (t) node[edgenode, pos=0.75, below=0.3cm] {$b_j$};
    
    \draw[line width=\thicc] (xi) -- (yj) node[edgenode, pos=0.5, above=0.2cm] {$1$};
    \draw[line width=\thicc] (xi) -- (yj') node[edgenode, pos=0.5, above=0.2cm] {};
    \draw[line width=\thicc] (xi') -- (yj) node[edgenode, pos=0.47, above=0.2cm] {$1$};
    
    \draw[line width=\thicc] (x'i) -- (y'j) node[edgenode, pos=0.5, above=0.2cm] {$1$};
    \draw[line width=\Thicc, dashed] (xi') -- (x'i) node[edgenode, pos=0.9, below left=0.1cm] {$a_i-1$};
    \draw[line width=\Thicc, dashed] (y'j) -- (yj') node[edgenode, pos=0.2, below right=0.1cm] {$b_j-1$};
    
    \begin{scope}[on background layer]
        \node[squarebox, rounded corners=10pt, fill=black, fill opacity=0.2, line width=\thicc, fit=(x1) (xi) (xl)] {};
    
        \node[squarebox, rounded corners=10pt, line width=\thicc, fit=(xl1) (xi') (xn)] {};
    
        \draw[black, fill=black, fill opacity=0.2, line width=\thicc]
            ($(x'1.north west)+(-0.56,0.2)$) [rounded corners=10pt] --
            ($(x'1.north east)+(0.56,0.2)$) [sharp corners] --
            ($(x'1.south east)+(0.56,0.7)$) [sharp corners] --
            ($(x'1.south west)+(-0.56,0.7)$) [rounded corners=10pt] -- cycle;
    
        \draw[black, line width=\thicc]
            ($(x'i.north west)+(-0.34,1.585)$) rectangle ($(x'i.south east)+(0.34,-0.585)$);
    
        \draw[black, fill=black, fill opacity=0.2, line width=\thicc]
            ($(x'n.north west)+(-0.56,0.9)$) [sharp corners] --
            ($(x'n.north east)+(0.56,0.9)$) [rounded corners=10pt] --
            ($(x'n.south east)+(0.56,-0.2)$) [rounded corners=10pt] --
            ($(x'n.south west)+(-0.56,-0.2)$) [sharp corners] -- cycle;
    
        \node[squarebox, rounded corners=10pt, line width=\thicc, fit=(y1) (yj) (yk)] {};
    
        \node[squarebox, rounded corners=10pt, fill=black, fill opacity=0.2, line width=\thicc, fit=(yk1) (yj') (ym)] {};
    
        \draw[black, fill=black, fill opacity=0.2, line width=\thicc]
            ($(y'1.north west)+(-0.56,0.2)$) [rounded corners=10pt] --
            ($(y'1.north east)+(0.56,0.2)$) [sharp corners] --
            ($(y'1.south east)+(0.56,-0.9)$) [sharp corners] --
            ($(y'1.south west)+(-0.56,-0.9)$) [rounded corners=10pt] -- cycle;
    
        \draw[black, line width=\thicc]
            ($(y'j.north west)+(-0.34,0.585)$) rectangle ($(y'j.south east)+(0.34,-0.985)$);
    
        \draw[black, fill=black, fill opacity=0.2, line width=\thicc]
            ($(y'm.north west)+(-0.56,-0.1)$) [sharp corners] --
            ($(y'm.north east)+(0.56,-0.1)$) [rounded corners=10pt] --
            ($(y'm.south east)+(0.56,-0.2)$) [rounded corners=10pt] --
            ($(y'm.south west)+(-0.56,-0.2)$) [sharp corners] -- cycle;
    \end{scope}
    
    \end{tikzpicture}
    }
    \caption{$s$-$t$ cut from $\calF_1$ in the flow network.
    }
    \label{fig:flow graph erdos-gallai 1}
    \end{subfigure}
    \hfill
    \begin{subfigure}{0.45\textwidth}
    \centering
    \resizebox{1\textwidth}{!}{
    \tikzset{
    myptr/.style={-{Stealth[scale=1.5]}},
    }
    
    \definecolor{r}{rgb}{1, 0, 0.0}
    \definecolor{g}{rgb}{0.0, 0.5, 0.0}
    \definecolor{b}{rgb}{0.0, 0, 1}
    \definecolor{rg}{rgb}{1, 0.5, 0.0}
    \definecolor{gb}{rgb}{0, 0.5, 1}
    \def\thicc{1.2}
    \def\Thicc{2.}
    
    \begin{tikzpicture}[
        roundnode/.style={circle, draw=black, minimum size=35pt, inner sep=0pt, line width=1pt, font=\Huge},
        squarebox/.style={draw=black, inner sep=5pt, font=\Huge},
        roundedbox/.style={draw=black, rounded corners=15pt, inner sep=5pt, font=\Huge}, 
        roundedboxprime/.style={draw=black,rounded corners=15pt, inner sep=5pt, font=\Huge}, %
        dots/.style={
        draw=none, fill=none, font=\Huge},
        edgenode/.style={font=\Huge}
    ]
    
    \node[dots] (x1) {$\vdots$};
    \node[roundnode, below=0.7cm of x1] (xi) {$x_{\iota}$};
    \node[dots, below=1.3cm of xi] (xl) {$\vdots$};
    
    \node[dots, below=2cm of xl] (xl1) {$\vdots$};
    \node[roundnode, below=1.3cm of xl1] (xi') {$x_i$};
    \node[dots, below=0.7cm of xi'] (xn) {$\vdots$};
    
    \node[font = \Huge, left=0.8cm of x1] (XD) {$X_D$};
    \node[font = \Huge, above=0.5cm of xl1] (XS) {$X_S$};
    
    \node[fit=(x1)(xn)] (xcenter) {};
    
    \node[roundnode, left=3cm of xcenter] (s) {$s$};
    
    \node[dots, below right=4cm of xl1] (x'1) {$\vdots$};
    \node[roundnode, below=1.3cm of x'1] (x'i) {$x'_i$};
    \node[dots, below=0.7cm of x'i] (x'n) {$\vdots$};
    
    \node[font = \Huge, above=0.5cm of x'1] (X'S) {$X'_S$};
    
    \node[dots, right=10cm of x1] (y1) {$\vdots$};
    \node[roundnode, below=1.3cm of y1] (yj) {$y_{\tau}$};
    \node[dots, below=0.7cm of yj] (yk) {$\vdots$};
    
    \node[dots, below=2cm of yk] (yk1) {$\vdots$};
    \node[roundnode, below=0.7cm of yk1] (yj') {$y_j$};
    \node[dots, below=1.3cm of yj'] (ym) {$\vdots$};

    \node[font = \Huge, right=0.8cm of y1] (YD) {$Y_D$};
    \node[font = \Huge, above=0.5cm of yk1] (YS) {$Y_S$};
    
    \node[fit=(y1)(ym)] (ycenter) {};
    
    \node[roundnode, fill=black!20, right=3cm of ycenter] (t) {$t$};
    
    \node[dots, below left=4cm of yk1] (y'1) {$\vdots$};
    \node[roundnode, below=0.7cm of y'1] (y'j) {$y'_j$};
    \node[dots, below=1.3cm of y'j] (y'm) {$\vdots$};
    
    \node[font = \Huge, above=0.5cm of y'1] (Y'S) {$Y'_S$};
    
    \draw[line width=\Thicc] (s) -- (xi) node[edgenode, pos=0.25, above=0.3cm] {$a_{\iota}$};
    \draw[line width=\Thicc] (s) -- (xi') node[edgenode, pos=0.25, below=0.3cm] {$a_i$};
    
    \draw[line width=\Thicc] (yj) -- (t) node[edgenode, pos=0.75, above=0.3cm] {$b_{\tau}$};
    \draw[line width=\Thicc] (yj') -- (t) node[edgenode, pos=0.75, below=0.3cm] {$b_j$};
    
    \draw[line width=\thicc] (xi) -- (yj) node[edgenode, pos=0.5, above=0.2cm] {$1$};
    \draw[line width=\thicc] (xi) -- (yj') node[edgenode, pos=0.5, above right=0.1cm] {$1$};
    \draw[line width=\thicc] (xi') -- (yj) node[edgenode, pos=0.5, above=0.2cm] {};
    
    \draw[line width=\thicc] (x'i) -- (y'j) node[edgenode, pos=0.5, above=0.2cm] {$1$};
    \draw[line width=\Thicc, dashed] (xi') -- (x'i) node[edgenode, pos=0.8, below left=0.1cm] {$a_i-1$};
    \draw[line width=\Thicc, dashed] (y'j) -- (yj') node[edgenode, pos=0.1, below right=0.1cm] {$b_j-1$};
    
    \begin{scope}[on background layer]
        \draw[black, fill=black, fill opacity=0.2, line width=\thicc] ($(x1.north west)+(-0.56,0.2)$) [rounded corners=10pt] -- ($(x1.north east)+(0.56,0.2)$) [sharp corners] -- ($(x1.south east)+(0.56,-0.6)$) [sharp corners] -- ($(x1.south west)+(-0.56,-0.6)$) [rounded corners=10pt] -- cycle;
         
        \draw[black, line width=\thicc] ($(xi.north west)+(-0.34,0.285)$)  rectangle ($(xi.south east)+(0.34,-0.285)$);
    
        \draw[black, fill=black, fill opacity=0.2, line width=\thicc] ($(xl.north west)+(-0.56,1.22)$) [sharp corners] -- ($(xl.north east)+(0.56,1.22)$) [rounded corners=10pt] -- ($(xl.south east)+(0.56,-0.2)$) [rounded corners=10pt]-- ($(xl.south west)+(-0.56,-0.2)$) [sharp corners] -- cycle;
    
        \draw[black, fill=black, fill opacity=0.2, line width=\thicc] ($(xl1.north west)+(-0.56,0.2)$) [rounded corners=10pt] -- ($(xl1.north east)+(0.56,0.2)$) [sharp corners] -- ($(xl1.south east)+(0.56,-0.9)$) [sharp corners] -- ($(xl1.south west)+(-0.56,-0.9)$) [rounded corners=10pt] -- cycle;
         
        \draw[black, line width=\thicc] ($(xi'.north west)+(-0.34,0.585)$)  rectangle ($(xi'.south east)+(0.34,-0.585)$);
    
        \draw[black, fill=black, fill opacity=0.2, line width=\thicc] ($(xn.north west)+(-0.56,0.3)$) [sharp corners] -- ($(xn.north east)+(0.56,0.3)$) [rounded corners=10pt] -- ($(xn.south east)+(0.56,-0.2)$) [rounded corners=10pt]-- ($(xn.south west)+(-0.56,-0.2)$) [sharp corners] -- cycle;
    
        \draw[black, fill=black, fill opacity=0.2, line width=\thicc]  ($(x'1.north west)+(-0.56,0.2)$) [rounded corners=10pt] -- ($(x'1.north east)+(0.56,0.2)$) [sharp corners] -- ($(x'1.south east)+(0.56,-0.9)$) [sharp corners] -- ($(x'1.south west)+(-0.56,-0.9)$) [rounded corners=10pt] -- cycle;
         
        \draw[black, line width=\thicc] ($(x'i.north west)+(-0.34,0.585)$)  rectangle ($(x'i.south east)+(0.34,-0.585)$);
    
        \draw[black, fill=black, fill opacity=0.2, line width=\thicc] ($(x'n.north west)+(-0.56,0.3)$) [sharp corners] -- ($(x'n.north east)+(0.56,0.3)$) [rounded corners=10pt] -- ($(x'n.south east)+(0.56,-0.2)$) [rounded corners=10pt]-- ($(x'n.south west)+(-0.56,-0.2)$) [sharp corners] -- cycle;
    
        \draw[black, fill=black, fill opacity=0.2, line width=\thicc] ($(y1.north west)+(-0.56,0.2)$) [rounded corners=10pt] -- ($(y1.north east)+(0.56,0.2)$) [sharp corners] -- ($(y1.south east)+(0.56,-0.6)$) [sharp corners] -- ($(y1.south west)+(-0.56,-0.6)$) [rounded corners=10pt] -- cycle;
         
        \draw[black, fill opacity=0.2, line width=\thicc] ($(yj.north west)+(-0.34,0.885)$)  rectangle ($(yj.south east)+(0.34,-0.885)$);
    
        \draw[black, fill=black, fill opacity=0.2, line width=\thicc]  ($(yk.north west)+(-0.56,0.0)$) [sharp corners] -- ($(yk.north east)+(0.56,0.0)$) [rounded corners=10pt] -- ($(yk.south east)+(0.56,-0.2)$) [rounded corners=10pt]-- ($(yk.south west)+(-0.56,-0.2)$) [sharp corners] -- cycle;
        
        \draw[black, fill=black, fill opacity=0.2, line width=\thicc]($(yk1.north west)+(-0.56,0.2)$) [rounded corners=10pt] -- ($(yk1.north east)+(0.56,0.2)$) [sharp corners] -- ($(yk1.south east)+(0.56,0.2)$) [sharp corners] -- ($(yk1.south west)+(-0.56,0.2)$) [rounded corners=10pt] -- cycle;
         
        \draw[black, line width=\thicc] ($(yj'.north west)+(-0.34,1.085)$)  rectangle ($(yj'.south east)+(0.34,-0.585)$);
    
        \draw[black, fill=black, fill opacity=0.2, line width=\thicc] ($(ym.north west)+(-0.56,0.9)$) [sharp corners] -- ($(ym.north east)+(0.56,0.9)$) [rounded corners=10pt] -- ($(ym.south east)+(0.56,-0.2)$) [rounded corners=10pt]-- ($(ym.south west)+(-0.56,-0.2)$) [sharp corners] -- cycle;
    
        \draw[black, fill=black, fill opacity=0.2, line width=\thicc] ($(y'1.north west)+(-0.56,0.2)$) [rounded corners=10pt] -- ($(y'1.north east)+(0.56,0.2)$) [sharp corners] -- ($(y'1.south east)+(0.56,0.2)$) [sharp corners] -- ($(y'1.south west)+(-0.56,0.2)$) [rounded corners=10pt] -- cycle;
         
        \draw[black, line width=\thicc] ($(y'j.north west)+(-0.34,1.085)$)  rectangle ($(y'j.south east)+(0.34,-0.585)$);
    
        \draw[black, fill=black, fill opacity=0.2, line width=\thicc] ($(y'm.north west)+(-0.56,0.9)$) [sharp corners] -- ($(y'm.north east)+(0.56,0.9)$) [rounded corners=10pt] -- ($(y'm.south east)+(0.56,-0.2)$) [rounded corners=10pt]-- ($(y'm.south west)+(-0.56,-0.2)$) [sharp corners] -- cycle;
    \end{scope}
    
    \end{tikzpicture}
    }
    \caption{$s$-$t$ cut from $\calF_2$ in the flow graph.}
    \label{fig:flow graph erdos-gallai 2}
    \end{subfigure}
    \par\medskip
    \begin{subfigure}{0.45\textwidth}
    \centering
    \resizebox{1\textwidth}{!}{
    \tikzset{
    myptr/.style={-{Stealth[scale=1.5]}},
    }
    
    \definecolor{r}{rgb}{1, 0, 0.0}
    \definecolor{g}{rgb}{0.0, 0.5, 0.0}
    \definecolor{b}{rgb}{0.0, 0, 1}
    \definecolor{rg}{rgb}{1, 0.5, 0.0}
    \definecolor{gb}{rgb}{0, 0.5, 1}
    \def\thicc{1.2}
    \def\Thicc{2.}
    
    \begin{tikzpicture}[
        roundnode/.style={circle, draw=black, minimum size=35pt, inner sep=0pt, line width=1pt, font=\Huge},
        squarebox/.style={draw=black, inner sep=5pt, font=\Huge},
        roundedbox/.style={draw=black, rounded corners=15pt, inner sep=5pt, font=\Huge}, 
        roundedboxprime/.style={draw=black,rounded corners=15pt, inner sep=5pt, font=\Huge}, %
        dots/.style={draw=none, fill=none, font=\Huge},
        edgenode/.style={font=\Huge}
    ]
    
    \node[dots] (x1) {$\vdots$};
    \node[roundnode, below=1cm of x1] (xi) {$x_{\iota}$};
    \node[dots, below=1cm of xi] (xl) {$\vdots$};
    
    \node[dots, below=2cm of xl] (xl1) {$\vdots$};
    \node[roundnode, below=1.3cm of xl1] (xi') {$x_i$};
    \node[dots, below=0.7cm of xi'] (xn) {$\vdots$};
    
    \node[font = \Huge, left=0.8cm of x1] (XD) {$X_D$};
    \node[font = \Huge, above=0.5cm of xl1] (XS) {$X_S$};
    
    \node[fit=(x1)(xn)] (xcenter) {};
    \node[roundnode, left=3cm of xcenter] (s) {$s$};
    
    \node[dots, below right=4cm of xl1] (x'1) {$\vdots$};
    \node[roundnode, below=1.3cm of x'1] (x'i) {$x'_i$};
    \node[dots, below=0.7cm of x'i] (x'n) {$\vdots$};
    \node[font = \Huge, above=0.5cm of x'1] (X'S) {$X'_S$};
    
    \node[dots, right=10cm of x1] (y1) {$\vdots$};
    \node[roundnode, below=1.3cm of y1] (yj) {$y_{\tau}$};
    \node[dots, below=0.7cm of yj] (yk) {$\vdots$};
    
    \node[dots, below=2cm of yk] (yk1) {$\vdots$};
    \node[roundnode, below=1cm of yk1] (yj') {$y_j$};
    \node[dots, below=1cm of yj'] (ym) {$\vdots$};
    
    \node[font = \Huge, right=0.8cm of y1] (YD) {$Y_D$};
    \node[font = \Huge, above=0.5cm of yk1] (YS) {$Y_S$};
    
    \node[fit=(y1)(ym)] (ycenter) {};
    \node[roundnode, fill=black!20, right=3cm of ycenter] (t) {$t$};
    
    \node[dots, below left=4cm of yk1] (y'1) {$\vdots$};
    \node[roundnode, below=0.7cm of y'1] (y'j) {$y'_j$};
    \node[dots, below=1.3cm of y'j] (y'm) {$\vdots$};
    \node[font = \Huge, above=0.5cm of y'1] (Y'S) {$Y'_S$};
    
    \draw[line width=\Thicc] (s) -- (xi) node[edgenode, pos=0.25, above=0.3cm] {$a_{\iota}$};
    \draw[line width=\Thicc] (s) -- (xi') node[edgenode, pos=0.25, below=0.3cm] {$a_i$};
    
    \draw[line width=\Thicc] (yj) -- (t) node[edgenode, pos=0.75, above=0.3cm] {$b_{\tau}$};
    \draw[line width=\Thicc] (yj') -- (t) node[edgenode, pos=0.75, below=0.3cm] {$b_j$};
    
    \draw[line width=\thicc] (xi) -- (yj) node[edgenode, pos=0.5, above=0.2cm] {$1$};
    \draw[line width=\thicc] (xi) -- (yj') node[edgenode, pos=0.5, above=0.2cm] {};
    \draw[line width=\thicc] (xi') -- (yj) node[edgenode, pos=0.52, above=0.3cm] {$1$};
    
    \draw[line width=\thicc] (x'i) -- (y'j) node[edgenode, pos=0.5, above=0.2cm] {$1$};
    \draw[line width=\Thicc, dashed] (xi') -- (x'i) node[edgenode, pos=0.8, below left=0.1cm] {$a_i-1$};
    \draw[line width=\Thicc, dashed] (y'j) -- (yj') node[edgenode, pos=0.1, below right=0.1cm] {$b_j-1$};
    
    \begin{scope}[on background layer]
        \node[squarebox, rounded corners=10pt, fill=black, fill opacity=0.2, line width=\thicc, fit=(x1) (xi) (xl)] {};
    
        \draw[black, fill=black, fill opacity=0.2, line width=\thicc]
            ($(xl1.north west)+(-0.56,0.5)$) [rounded corners=10pt] --
            ($(xl1.north east)+(0.56,0.5)$) [sharp corners] --
            ($(xl1.south east)+(0.56,-0.9)$) [sharp corners] --
            ($(xl1.south west)+(-0.56,-0.9)$) [rounded corners=10pt] -- cycle;
    
        \draw[black, line width=\thicc]
            ($(xi'.north west)+(-0.34,0.585)$) rectangle ($(xi'.south east)+(0.34,-1.085)$);
    
        \draw[black, fill=black, fill opacity=0.2, line width=\thicc]
            ($(xn.north west)+(-0.56,-0.2)$) [sharp corners] --
            ($(xn.north east)+(0.56,-0.2)$) [rounded corners=10pt] --
            ($(xn.south east)+(0.56,-0.2)$) [rounded corners=10pt] --
            ($(xn.south west)+(-0.56,-0.2)$) [sharp corners] -- cycle;
    
        \draw[black, fill=black, fill opacity=0.2, line width=\thicc]
            ($(x'1.north west)+(-0.56,0.2)$) [rounded corners=10pt] --
            ($(x'1.north east)+(0.56,0.2)$) [sharp corners] --
            ($(x'1.south east)+(0.56,-0.9)$) [sharp corners] --
            ($(x'1.south west)+(-0.56,-0.9)$) [rounded corners=10pt] -- cycle;
    
        \draw[black, line width=\thicc]
            ($(x'i.north west)+(-0.34,0.585)$) rectangle ($(x'i.south east)+(0.34,-1.085)$);
    
        \draw[black, fill=black, fill opacity=0.2, line width=\thicc]
            ($(x'n.north west)+(-0.56,-0.2)$) [sharp corners] --
            ($(x'n.north east)+(0.56,-0.2)$) [rounded corners=10pt] --
            ($(x'n.south east)+(0.56,-0.2)$) [rounded corners=10pt] --
            ($(x'n.south west)+(-0.56,-0.2)$) [sharp corners] -- cycle;
    
        \draw[black, fill=black, fill opacity=0.2, line width=\thicc]
            ($(y1.north west)+(-0.56,0.2)$) [rounded corners=10pt] --
            ($(y1.north east)+(0.56,0.2)$) [sharp corners] --
            ($(y1.south east)+(0.56,-0.9)$) [sharp corners] --
            ($(y1.south west)+(-0.56,-0.9)$) [rounded corners=10pt] -- cycle;
    
        \draw[black, line width=\thicc]
            ($(yj.north west)+(-0.34,0.585)$) rectangle ($(yj.south east)+(0.34,-0.285)$);
    
        \draw[black, fill=black, fill opacity=0.2, line width=\thicc]
            ($(yk.north west)+(-0.56,0.6)$) [sharp corners] --
            ($(yk.north east)+(0.56,0.6)$) [rounded corners=10pt] --
            ($(yk.south east)+(0.56,-0.2)$) [rounded corners=10pt] --
            ($(yk.south west)+(-0.56,-0.2)$) [sharp corners] -- cycle;
    
        \node[squarebox, rounded corners=10pt, fill=black, fill opacity=0.2, line width=\thicc, fit=(yk1) (yj') (ym)] {};
    
        \draw[black, fill=black, fill opacity=0.2, line width=\thicc]
            ($(y'1.north west)+(-0.56,0.2)$) [rounded corners=10pt] --
            ($(y'1.north east)+(0.56,0.2)$) [sharp corners] --
            ($(y'1.south east)+(0.56,0.1)$) [sharp corners] --
            ($(y'1.south west)+(-0.56,0.1)$) [rounded corners=10pt] -- cycle;
    
        \draw[black, line width=\thicc]
            ($(y'j.north west)+(-0.34,0.985)$) rectangle ($(y'j.south east)+(0.34,-0.585)$);
    
        \draw[black, fill=black, fill opacity=0.2, line width=\thicc]
            ($(y'm.north west)+(-0.56,0.9)$) [sharp corners] --
            ($(y'm.north east)+(0.56,0.9)$) [rounded corners=10pt] --
            ($(y'm.south east)+(0.56,-0.2)$) [rounded corners=10pt] --
            ($(y'm.south west)+(-0.56,-0.2)$) [sharp corners] -- cycle;
    \end{scope}
    
    \end{tikzpicture}
    }
    \caption{$s$-$t$ cut from $\calF_3$ in the flow graph.}
    \label{fig:flow graph erdos-gallai 3}
    \end{subfigure}
    \hfill
    \begin{subfigure}{0.45\textwidth}
    \centering
    \resizebox{1\textwidth}{!}{
    \tikzset{
    myptr/.style={-{Stealth[scale=1.5]}},
    }
    
    \definecolor{r}{rgb}{1, 0, 0.0}
    \definecolor{g}{rgb}{0.0, 0.5, 0.0}
    \definecolor{b}{rgb}{0.0, 0, 1}
    \definecolor{rg}{rgb}{1, 0.5, 0.0}
    \definecolor{gb}{rgb}{0, 0.5, 1}
    \def\thicc{1.2}
    \def\Thicc{2.0}
    
    \begin{tikzpicture}[
        roundnode/.style={circle, draw=black, minimum size=35pt, inner sep=0pt, line width=1pt, font=\Huge},
        squarebox/.style={draw=black, inner sep=5pt, font=\Huge},
        roundedbox/.style={draw=black, rounded corners=15pt, inner sep=5pt, font=\Huge}, 
        roundedboxprime/.style={draw=black,rounded corners=15pt, inner sep=5pt, font=\Huge}, %
        dots/.style={draw=none, fill=none, font=\Huge},
        edgenode/.style={font=\Huge}
    ]
    
    \node[dots] (x1) {$\vdots$};
    \node[roundnode, below=1cm of x1] (xi) {$x_{\iota}$};
    \node[dots, below=1cm of xi] (xl) {$\vdots$};
    
    \node[dots, below=2cm of xl] (xl1) {$\vdots$};
    \node[roundnode, below=1.3cm of xl1] (xi') {$x_i$};
    \node[dots, below=0.7cm of xi'] (xn) {$\vdots$};
    
    \node[font = \Huge, left=0.8cm of x1] (XD) {$X_D$};
    \node[font = \Huge, above=0.5cm of xl1] (XS) {$X_S$};
    
    \node[fit=(x1)(xn)] (xcenter) {};
    \node[roundnode, left=3cm of xcenter] (s) {$s$};
    
    \node[dots, below right=4cm of xl1] (x'1) {$\vdots$};
    \node[roundnode, below=1.3cm of x'1] (x'i) {$x'_i$};
    \node[dots, below=0.7cm of x'i] (x'n) {$\vdots$};
    
    \node[font = \Huge, above=0.5cm of x'1] (X'S) {$X'_S$};
    
    \node[dots, right=10cm of x1] (y1) {$\vdots$};
    \node[roundnode, below=1.3cm of y1] (yj) {$y_{\tau}$};
    \node[dots, below=0.7cm of yj] (yk) {$\vdots$};
    
    \node[dots, below=2cm of yk] (yk1) {$\vdots$};
    \node[roundnode, below=1cm of yk1] (yj') {$y_j$};
    \node[dots, below=1cm of yj'] (ym) {$\vdots$};
    
    \node[font = \Huge, right=0.8cm of y1] (YD) {$Y_D$};
    \node[font = \Huge, above=0.5cm of yk1] (YS) {$Y_S$};
    
    \node[fit=(y1)(ym)] (ycenter) {};
    \node[roundnode, fill=black!20, right=3cm of ycenter] (t) {$t$};
    
    \node[dots, below left=4cm of yk1] (y'1) {$\vdots$};
    \node[roundnode, below=0.7cm of y'1] (y'j) {$y'_j$};
    \node[dots, below=1.3cm of y'j] (y'm) {$\vdots$};
    
    \node[font = \Huge, above=0.5cm of y'1] (Y'S) {$Y'_S$};
    
    \draw[line width=\Thicc] (s) -- (xi) node[edgenode, pos=0.25, above=0.3cm] {$a_{\iota}$};
    \draw[line width=\Thicc] (s) -- (xi') node[edgenode, pos=0.25, below=0.3cm] {$a_i$};
    
    \draw[line width=\Thicc] (yj) -- (t) node[edgenode, pos=0.75, above=0.3cm] {$b_{\tau}$};
    \draw[line width=\Thicc] (yj') -- (t) node[edgenode, pos=0.75, below=0.3cm] {$b_j$};
    
    \draw[line width=\thicc] (xi) -- (yj) node[edgenode, pos=0.5, above=0.2cm] {$1$};
    \draw[line width=\thicc] (xi) -- (yj') node[edgenode, pos=0.5, above=0.2cm] {};
    \draw[line width=\thicc] (xi') -- (yj) node[edgenode, pos=0.52, above=0.3cm] {$1$};
    
    \draw[line width=\thicc] (x'i) -- (y'j) node[edgenode, pos=0.5, above=0.2cm] {$1$};
    \draw[line width=\Thicc, dashed] (xi') -- (x'i) node[edgenode, pos=0.8, below left=0.1cm] {$a_i-1$};
    \draw[line width=\Thicc, dashed] (y'j) -- (yj') node[edgenode, pos=0.1, below right=0.1cm] {$b_j-1$};
    
    \begin{scope}[on background layer]
        \draw[black, fill=black, fill opacity=0.2, line width=\thicc]
            ($(x1.north west)+(-0.56,0.2)$) [rounded corners=10pt] --
            ($(x1.north east)+(0.56,0.2)$) [sharp corners] --
            ($(x1.south east)+(0.56,-0.6)$) [sharp corners] --
            ($(x1.south west)+(-0.56,-0.6)$) [rounded corners=10pt] -- cycle;
    
        \draw[black, line width=\thicc]
            ($(xi.north west)+(-0.34,0.585)$) rectangle ($(xi.south east)+(0.34,0.05)$);
    
        \draw[black, fill=black, fill opacity=0.2, line width=\thicc]
            ($(xl.north west)+(-0.56,1.22)$) [sharp corners] --
            ($(xl.north east)+(0.56,1.22)$) [rounded corners=10pt] --
            ($(xl.south east)+(0.56,-0.2)$) [rounded corners=10pt] --
            ($(xl.south west)+(-0.56,-0.2)$) [sharp corners] -- cycle;
    
        \node[squarebox, rounded corners=10pt, line width=\thicc, fit=(xl1) (xi') (xn)] {};
    
        \draw[black, fill=black, fill opacity=0.2, line width=\thicc]
            ($(x'1.north west)+(-0.56,0.2)$) [rounded corners=10pt] --
            ($(x'1.north east)+(0.56,0.2)$) [sharp corners] --
            ($(x'1.south east)+(0.56,-0.9)$) [sharp corners] --
            ($(x'1.south west)+(-0.56,-0.9)$) [rounded corners=10pt] -- cycle;
    
        \draw[black, line width=\thicc]
            ($(x'i.north west)+(-0.34,0.585)$) rectangle ($(x'i.south east)+(0.34,-1.085)$);
    
        \draw[black, fill=black, fill opacity=0.2, line width=\thicc]
            ($(x'n.north west)+(-0.56,-0.2)$) [sharp corners] --
            ($(x'n.north east)+(0.56,-0.2)$) [rounded corners=10pt] --
            ($(x'n.south east)+(0.56,-0.2)$) [rounded corners=10pt] --
            ($(x'n.south west)+(-0.56,-0.2)$) [sharp corners] -- cycle;
    
        \node[squarebox, rounded corners=10pt, line width=\thicc, fit=(y1) (yj) (yk)] {};
    
        \draw[black, fill=black, fill opacity=0.2, line width=\thicc]
            ($(yk1.north west)+(-0.56,0.2)$) [rounded corners=10pt] --
            ($(yk1.north east)+(0.56,0.2)$) [sharp corners] --
            ($(yk1.south east)+(0.56,0.2)$) [sharp corners] --
            ($(yk1.south west)+(-0.56,0.2)$) [rounded corners=10pt] -- cycle;
    
        \draw[black, line width=\thicc]
            ($(yj'.north west)+(-0.34,1.385)$) rectangle ($(yj'.south east)+(0.34,-0.285)$);
    
        \draw[black, fill=black, fill opacity=0.2, line width=\thicc]
            ($(ym.north west)+(-0.56,0.9)$) [sharp corners] --
            ($(ym.north east)+(0.56,0.9)$) [rounded corners=10pt] --
            ($(ym.south east)+(0.56,-0.2)$) [rounded corners=10pt] --
            ($(ym.south west)+(-0.56,-0.2)$) [sharp corners] -- cycle;
    
        \draw[black, fill=black, fill opacity=0.2, line width=\thicc]
            ($(y'1.north west)+(-0.56,0.2)$) [rounded corners=10pt] --
            ($(y'1.north east)+(0.56,0.2)$) [sharp corners] --
            ($(y'1.south east)+(0.56,0.2)$) [sharp corners] --
            ($(y'1.south west)+(-0.56,0.2)$) [rounded corners=10pt] -- cycle;
    
        \draw[black, line width=\thicc]
            ($(y'j.north west)+(-0.34,1.085)$) rectangle ($(y'j.south east)+(0.34,-0.585)$);
    
        \draw[black, fill=black, fill opacity=0.2, line width=\thicc]
            ($(y'm.north west)+(-0.56,0.9)$) [sharp corners] --
            ($(y'm.north east)+(0.56,0.9)$) [rounded corners=10pt] --
            ($(y'm.south east)+(0.56,-0.2)$) [rounded corners=10pt] --
            ($(y'm.south west)+(-0.56,-0.2)$) [sharp corners] -- cycle;
    \end{scope}
    
    \end{tikzpicture}
    }
    \caption{$s$-$t$ cut from $\calF_4$ in the flow graph.}
    \label{fig:flow graph erdos-gallai 4}
    \end{subfigure}
    \caption{Schematic depiction of the cut families. The white areas represent $S$, while the gray areas represent $T$.}
    \label{fig:flow graph erdos-gallai}
\end{figure}
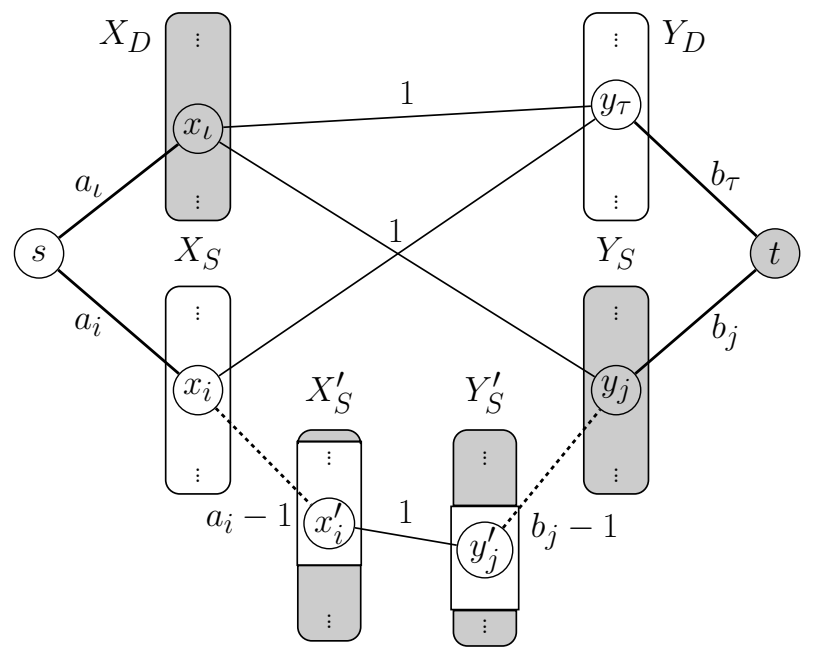
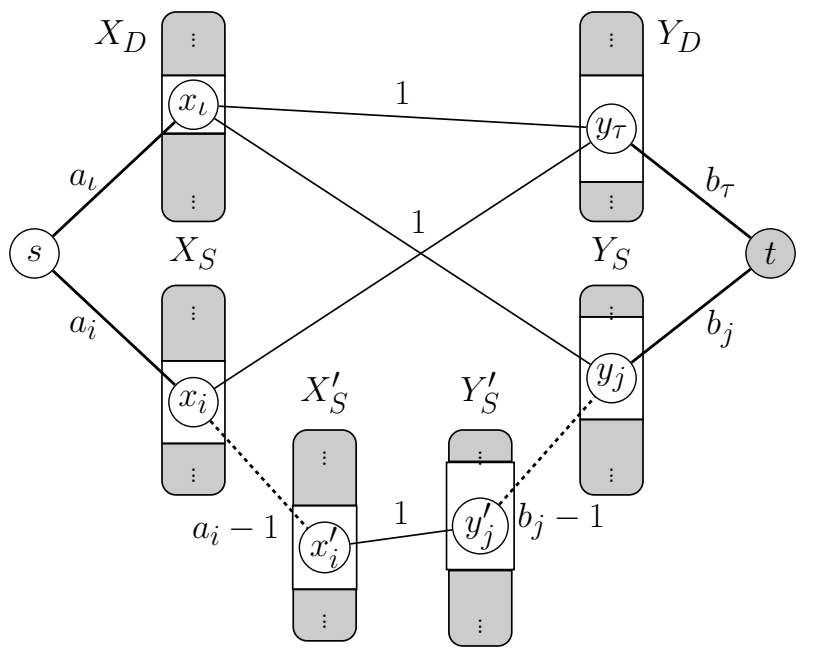
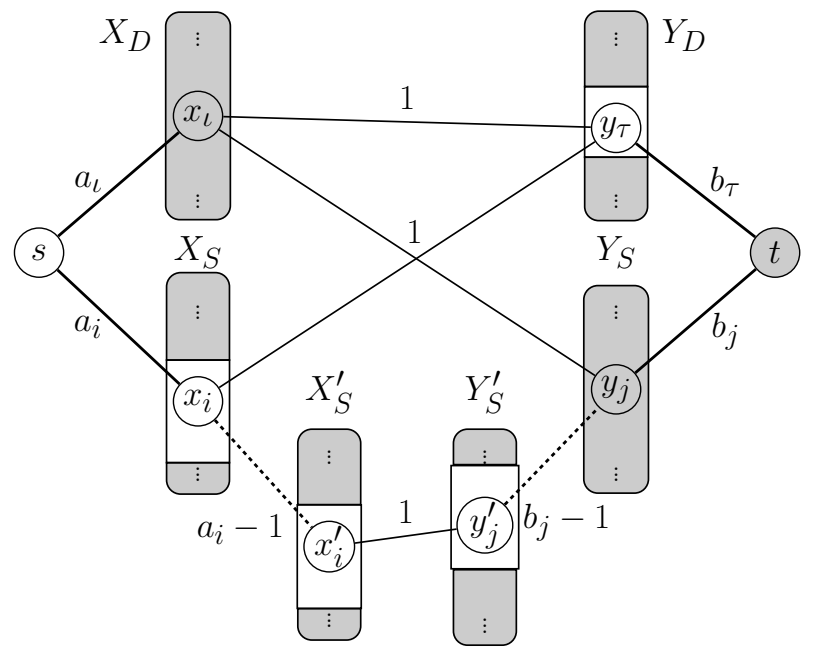
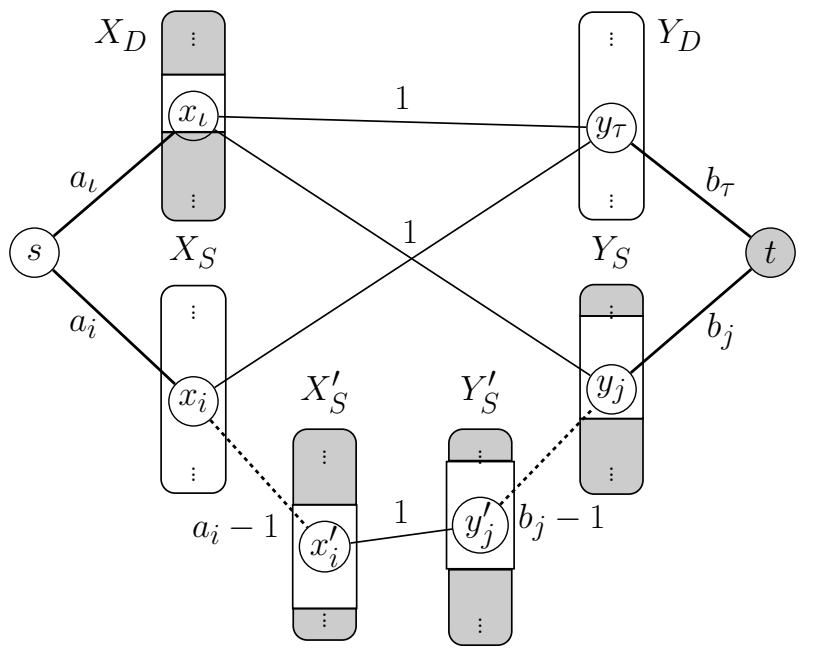
} %

\begin{credits}
\subsubsection{\discintname}
The authors have no competing interests to declare that are
relevant to the content of this article.
\end{credits}


\bibliographystyle{splncs04.bst}
\bibliography{realizations-pc}

@string{DAM = "Discr. Appl. Math."}

@string{DM = "Discr. Math."}

@string{LNCS = "LNCS"}

@inproceedings{BKPR26-domset,
  author    = {Amotz Bar-Noy and Igor Kalinichev and David Peleg and Dror Rawitz},
  title     = {Degree Realization with Minimum Dominating Set},
  booktitle = {27th IPCO},
  series       = LNCS,
  volume       = {16588},
  pages        = {64--80},
  year         = {2026},
}

@inproceedings{BKPR26-matching,
  author    = {Amotz Bar-Noy and Igor Kalinichev and David Peleg and Dror Rawitz},
  title     = {Degree Realization with Maximum Matching},
  booktitle = {37th IWOCA},
  series       = LNCS,
  volume       = {16587},
  pages        = {103--117},
  year         = {2026},
}

@article{BockRautenbach19,
  author       = {Felix Bock and Dieter Rautenbach},
  title        = {On matching numbers of tree and bipartite degree sequences},
  doi          = {10.1016/j.disc.2019.02.005},
  journal      = DM,
  volume       = {342},
  number       = {6},
  pages        = {1687--1695},
  year         = {2019},
}

@Article{EG60,
  author = 	 {Paul Erd\H{o}s and Tibor Gallai},
  title = 	 {Graphs with Prescribed Degrees of Vertices {[Hungarian]}},
  journal = 	 {Matematikai Lapok},
  year = 	 {1960},
  volume = 	 {11},
  pages = 	 {264--274},
}

@article{EKMT24,
   title={New Results on Graph Matching from Degree-Preserving Growth},
   author={Erd{\H{o}}s, Péter L. and Kharel, Shubha R. and Mezei, Tamás Róbert and Toroczkai, Zoltán},
   doi ={10.3390/math12223518},
   volume={12},
   number={22},
   pages={3518},
   journal={Mathematics},
   year={2024},
}

@article{FHM1965,
    title={Some Properties of Graphs with Multiple Edges},
    author={D. R. Fulkerson and A. J. Hoffman and M. H. McAndrew},
    journal={Canadian Journal of Mathematics},
    volume={17},
    pages={166--177},
    year={1965},
}

@article{Gale57,
author    = {D. Gale},
title     = {A theorem on flows in networks},
journal   = {Pacific J. Math.},
volume    = {7},
pages     = {1073--1082},
year      = {1957}
}

@article{GHR16,
  author       = {Michael Gentner and Michael A. Henning and Dieter Rautenbach},
  title        = {Largest domination number and smallest independence number of forests
                  with given degree sequence},
  doi          = {10.1016/j.dam.2016.01.040},
  journal      = DAM,
  volume       = {206},
  pages        = {181--187},
  year         = {2016},
}

@article{GHR18,
  author       = {Michael Gentner and Michael A. Henning and Dieter Rautenbach},
  title        = {Smallest domination number and largest independence number of graphs
                  and forests with given degree sequence},
  doi          = {10.1002/jgt.22189},
  journal      = {J. Graph Theory},
  volume       = {88},
  number       = {1},
  pages        = {131--145},
  year         = {2018},
}

@inproceedings{GJL99,
  author    = {R. J. Gould and M. S. Jackson and J. Lehel},
  title     = {Potentially $G$-graphical degree sequences},
  booktitle = {Combinatorics, Graph Theory, and Algorithms},
  volume = {1},
  pages     = {451--460},
  year      = {1999},
}

@article{Hakimi62,
    author={Seifollah Louis Hakimi},
    title={On Realizability of a Set of Integers as Degrees of the Vertices of a Linear Graph -- {I}},
    journal = {SIAM J. Appl. Math.},
    volume={10},
    number = {3},
    pages={496--506},
    year={1962},
    publisher={Society for Industrial and Applied Mathematics}
}

@Article{havel55-long,
  author=      {Havel, V\'{a}clav},
  title = 	 {A Remark on the Existence of Finite Graphs [in {C}zech]},
  journal =    {Časopis pro pěstování matematiky},
  year = 	     {1955},
  volume = 	 {80},
  pages = 	 {477--480},
}

@article{Konig31,
  author       = {K{\H{o}}nig, Dénes},
  title        = {Gr\'{a}fok \'{e}s m\'{a}trixok},
  journal      = {Matematikai \'{e}s Fizikai Lapok},
  volume       = {38},
  pages        = {116--119},
  year         = {1931},
}

@inproceedings{Orlin13,
    author = {Orlin, James B.},
    title = {Max flows in {$O(nm)$} time, or better},
    year = {2013},
    isbn = {9781450320290},
    publisher = {Association for Computing Machinery},
    doi = {10.1145/2488608.2488705},
    booktitle = {Proceedings of the Forty-Fifth Annual ACM Symposium on Theory of Computing},
    pages = {765–774},
    numpages = {10},
    series = {STOC '13}
}

@inproceedings{Rao79,
author    = {A. R. Rao},
title     = {The clique number of a graph with a given degree sequence},
Booktitle = {ISI Lect. Notes},
year      = {1979},
volume    = {4},
pages     = {251--267},
}

@article{Ryser57,
author    = {H.J. Ryser},
title     = {Combinatorial properties of matrices of zeros and ones},
journal   = {Canad. J. Math.},
volume    = {9},
pages     = {371--377},
year      = {1957}
}

@article{Yin12,
  author       = {Jian-Hua Yin},
  title        = {A short constructive proof of {A. R. Rao}'s characterization of potentially
                    {$K_{r+1}$-graphic} sequences},
  doi       = {10.1016/j.dam.2011.10.015},
  journal      = DAM,
  volume       = {160},
  number       = {3},
  pages        = {352--354},
  year         = {2012},
}

\clearpage

\appendix
\centerline{\bf\LARGE Appendix}

\bigskip
\bigskip
\bigskip
\bigskip
\section{Proofs of the characterization theorem}
\label{sec:proofs}

\APPENDA

\begin{lemma}
\label{lemma:modification emptyY}
Let $(S, T)$ be a cut of $G_{a, b, Q}$ such that $T \cap Y_D = \emptyset$.
Also, let $S' = S \cup X_S$ and $T' = T \setminus X_S$.
Then $c(S',T') \le c(S,T)$.
\end{lemma}
\begin{proof}
Let $x_i \in X_S \setminus S$.
By adding $x_i$ to $S'$ we remove the edge $(s,x_i)$ from the cut, but we may add 
the edge $(x_i,x'_i)$ to the cut (see \Cref{fig:flow graph}). Hence, 
\[
c(S,T) - c(S', T') \geq \sum_{x_i \in X_S \setminus S}(a_i - (a_i-1)) \ge 0
~.
\]
\qed\end{proof}

Symmetric arguments imply that 

\begin{lemma}
\label{lemma:modification emptyX}
Let $(S,T)$ be a cut of $G_{a, b, Q}$ such that $S \cap X_D = \emptyset$.
Also, let $T' = T \cup Y_S$ and $S' = S \setminus Y_S$.
Then $c(S',T') \le c(S,T)$.
\end{lemma}

\begin{lemma}
\label{lemma:modification non-emptyY}
Let $(S, T)$ be a cut of $G_{a, b, Q}$ such that $T \cap Y_D \neq \emptyset$.
Also, let $S' = S \setminus(Z \cup Z')$ and $T'= T \cup (Z \cup Z')$, where 
$Z  = \{x_i \in S \cap X_S\mid x'_i \notin S\}$ and 
$Z' =  \{x'_i \in S \cap X'_S\mid x_i \notin S\}$.
Then $c(S',T') \le c(S,T)$.
\end{lemma}
\begin{proof}
Let $x_i \in Z$,
 removal of $x_i$ adds the edge $(s,x_i)$ to the cut, but removes the edge $(x_i,x'_i)$ and at least 
one edge of the form $(x_i,y_j)$, where $y_j \in Y_D$.
Next, let $x'_i \in Z'$,
the removal of $x'_i$ removes edges of the form $(x'_i,y'_j)$,
where $y'_j \in Y'_S \cap T$.
(See \Cref{fig:flow graph}.)
Since $T \cap Y_D \neq \emptyset$, it follows that 
\[
c(S,T) - c(S', T') 
= \sum_{x_i \in Z} (a_i-1 + |T \cap Y_D| - a_i) + \sum_{x'_i \in Z'} |T \cap Y'_S| 
\geq 0
~. 
\]
\qed\end{proof}

Symmetric arguments imply that 

\begin{lemma}
\label{lemma:modification non-emptyX}
Let $(S, T)$ be a cut of $G_{a, b, Q}$ such that $S \cap X_D \neq \emptyset$.
Also, let $S' = S \cup (Z \cup Z')$ and $T'= T \setminus (Z \cup Z')$, where 
$Z  = \{y_i \in T \cap Y_S \mid y'_i \notin T\}$ and $Z' =  \{y'_i \in T \cap Y'_S \mid y_i \notin T\}$.
Then $c(S',T') \le c(S,T)$.
\end{lemma}

\inline Proof of Lemma \ref{cor:one}: 

Let $(S,T)$ be an arbitrary $s$-$t$ cut in $G_{a, b, Q}$. 
We consider the following four cases:
\begin{description}%
\item[Case 1:] If $(S,T) \in \calS_{00}$, then by \Cref{lemma:modification emptyY,lemma:modification emptyX}, 
      there exists a cut $(S',T') \in \calF_{1}$ such that $c(S', T') \leq c(S,T)$.

\item[Case 2:] If $(S,T) \in \calS_{11}$, then by \Cref{lemma:modification non-emptyY,lemma:modification non-emptyX}, 
      there exists a cut $(S',T') \in \calF_2$ such that $c(S', T') \leq c(S,T)$.

\item[Case 3:] If $(S,T) \in \calS_{01}$, then by \Cref{lemma:modification non-emptyY,lemma:modification emptyX}, 
      there exists a cut $(S',T') \in \calF_3$ such that $c(S', T') \leq c(S,T)$.

\item[Case 4:] If $(S,T) \in \calS_{10}$, then by \Cref{lemma:modification emptyY,lemma:modification non-emptyX}, 
      there exists a cut $(S',T') \in \calF_4$ such that $c(S', T') \leq c(S,T)$.
\hfill$\Box$
\end{description}

\inline Proof of Lemma \ref{lemma:F1-F4}: 

We prove each of the four parts of the lemma separately.

\begin{lemma}
\label{lemma:F1}
$c(S,T) \geq \sum_{i=1}^n a_i$, for every $(S,T) \in \calF_1$ if and only if 
the System~\eqref{eqn:erdos-gallai 1} is satisfied.
\end{lemma}
\begin{proof}
Consider a cut $(S^1_{I,J}, V\setminus S^1_{I,J}) \in \calF_1$, 
where $I \subseteq \IXS$, $J \subseteq \IYS$.
We have
\[
c(S^1_{I,J}, V\setminus S^1_{I,J}) 
= \sum_{i\in \IXD} a_i ~+~ \sum_{i\in\IXS\setminus I} (a_i-1) ~+~
  \sum_{i \in I, j \in \IYS \setminus J} 1 ~+~ 
  \sum_{j \in J} (b_j-1) ~+~ \sum_{j\in \IYD} b_j
~.
\]
(See an example in~\Cref{fig:flow graph erdos-gallai 1}.)
Hence capacity of all cuts in $\calF_1$ is least $\sum_{i=1}^n a_i$ if and only if 
for every $I \subseteq \IXS$, $J \subseteq \IYS$ we have
\[
\sum_{i\in I}a_i 
\leq \sum_{j\in \IYD} b_j - (|I_S| -|I|) + 
     \sum_{j \in \IYS \setminus J} |I| + 
     \sum_{j \in J}(b_j-1) 
~.
\]
It follows that this system is satisfied if and only if 
for every prefix $I \subseteq \IXS$
\[
\sum_{i\in I} (a_i-1) 
\leq \sum_{j\in \IYD} b_j - |I_S| + 
     \sum_{j \in \IYS} \min\{|I|, b_j-1\}
~.
\]
\qed\end{proof}

\begin{lemma}
\label{lemma:F2}
$c(S,T) \geq \sum_{i=1}^n a_i$, for every $(S,T) \in \calF_2$ if and only if 
the System~\eqref{eqn:erdos-gallai 2} is satisfied.
\end{lemma}
\begin{proof}
Consider a cut $(S^2_{I,J}, V\setminus S^2_{I,J}) \in \calF_2$, 
where $I \subseteq [1,n]$ and $J \subseteq [1,m]$. 
Observe that
\begin{align*}
c(S^2_{I,J}, V\setminus S^2_{I,J}) 
& = \sum_{i \in [1,n] \setminus I} a_i + 
    \sum_{i \in I, j \in [1,m] \setminus J} 1 +
    \sum_{j \in J} b_j~.
\end{align*}
(See an example in~\Cref{fig:flow graph erdos-gallai 2}.)
It follows that the capacity of all cuts in $\calF_2$ is at least $\sum_{i=1}^n a_i$ if and only if for every $I\subseteq [1,n]$ and $J \subseteq [1,m]$, we have
\[
\sum_{i\in I} a_i \le \sum_{j \in [1,m] \setminus J} |I| + \sum_{j \in J} b_j
~.
\]
It follows that this system is satisfied if and only if for every prefix $I \subseteq[1,n]$
\[
\sum_{i\in I} a_i \le \sum_{j \in [1,m]} \min\{|I|, b_j\}
~.
\]
\qed\end{proof}

\begin{lemma}
\label{lemma:F3}
$c(S,T) \geq \sum_{i=1}^n a_i$, for every $(S,T) \in \calF_3$ if and only if the System~\eqref{eqn:erdos-gallai 3} is satisfied.
\end{lemma}
\begin{proof}
Consider a cut $(S^3_{I,J}, V\setminus S^3_{I,J}) \in \calF_3$, where $I \subseteq \IXS$ 
and $J \subseteq [1,m]$.
Observe that
\[
c(S^3_{I,J}, V\setminus S^3_{I,J}) 
= \sum_{i\in \IXD} a_i ~+~ \sum_{i\in \IXS \setminus I} a_i ~+~ 
  \sum_{i \in I, j \in [1,m] \setminus J} 1 ~+~ 
  \sum_{j\in \IYD \cap  J} b_j ~+~ \sum_{j \in \IYS \cap J} (b_j-1) 
~. 
\]
(See an example in~\Cref{fig:flow graph erdos-gallai 3}.)
It follows that the capacity of all cuts in $\calF_3$ is at least $\sum_{i=1}^n a_i$ if and only if for every $I \subseteq \IXS$ and $J \subseteq [1,m]$
we have
\[
\sum_{i\in I}a_i 
\le \sum_{j \in [1,m] \setminus J} |I|  + 
    \sum_{j \in \IYD\cap  J} b_j + \sum_{j \in \IYS \cap J} (b_j-1) 
    ~.
\]
It follows that this system is satisfied if and only if for every prefix
$I \subseteq \IXS$
\[
\sum_{i\in I} a_i 
\leq \sum_{j \in \IYD} \min\{|I|, b_j\} + \sum_{j\in \IYS} \min\{|I|, b_j-1\}
~.
\]
\qed\end{proof}

\begin{lemma}
\label{lemma:F4}
$c(S,T) \geq \sum_{i=1}^n a_i$, for every $(S,T) \in \calF_4$ if and only if the System~\eqref{eqn:erdos-gallai 4} is satisfied.
\end{lemma}
\begin{proof}
Consider a cut $(S^4_{I,J}, V\setminus S^4_{I,J}) \in \calF_4$, where $I \subseteq [1,n]$ 
and $J \subseteq \IYS$.
Observe that
\begin{align*}
c(S^4_{I,J}, V\setminus S^4_{I,J}) 
& = \sum_{i\in \IXD \setminus I} a_i + \sum_{i\in \IXS \setminus I} (a_i-1)  + 
    \sum_{i \in I, j \in \IYS \setminus J} 1 + \sum_{j\in \IYD} b_j +\sum_{j \in J} b_j ~. 
\end{align*}
(See an example in~\Cref{fig:flow graph erdos-gallai 4}.)
It follows that the capacity of all cuts in $\calF_4$ is at least 
$\sum_{i=1}^n a_i = \sum_{j=1}^m b_j$ 
if and only if for every $I \subseteq [1,n]$ and $J \subseteq \IYS$, we have
\[
\sum_{j\in \IYS \setminus J}b_j 
\le \sum_{i \in I} |\IYS \setminus J|  + \sum_{i\in \IXD \setminus I} a_i + \sum_{i\in \IXS \setminus I} (a_i-1)
    ~.
\]
It is not hard to verify that this system is satisfied if and only if for every prefix
$J \subseteq \IYS$
\[
\sum_{j\in J} b_j 
\leq \sum_{i \in \IXD} \min\{|J|, a_i\} + \sum_{i \in \IXS} \min\{|J|, a_i-1\}
~. 
\]
\qed\end{proof}
\section{Complexity}
Any part of a bipartite graph is a dominating set in any realization. Hence, the optimal dominating set size is at most $\min(n,m)$. Using the characterization given by Theorem \ref{thm:systems} and a binary search over sizes of a dominating set we can find its minimum size in time $O(nm(n+m)\min(n,m)\log(\min(n,m))) = O(n^2m^2 \log(\min(n,m)))$. Moreover, the characterizations also give optimal pair of intervals $I_D$, $J_D$ for the dominating set. Finally, applying the Orlin’s maximum flow algorithm to the network corresponding to this pair of intervals yields the desired realization in $O(nm(n+m))$ additional time, which does not increase the overall asymptotic running time. 
\end{document}